\documentclass[fleqn,usenatbib]{mnras}

\usepackage{newtxtext,newtxmath}

\usepackage[T1]{fontenc}
\usepackage[dvipsnames]{xcolor}
\usepackage{makecell}
\usepackage{longtable}
\usepackage{comment}

\DeclareUnicodeCharacter{2212}{-}
\usepackage[utf8]{inputenc}

\inputencoding{utf8}
\DeclareRobustCommand{\VAN}[3]{#2}
\let\VANthebibliography\thebibliography
\def\thebibliography{\DeclareRobustCommand{\VAN}[3]{##3}\VANthebibliography}

\usepackage{graphicx}	
\usepackage{amsmath}	
\usepackage{multicol}

\usepackage{caption, subcaption}
\usepackage[normalem]{ulem}
\useunder{\uline}{\ul}{}

\usepackage[para,online,flushleft]{threeparttable}
\usepackage{hyperref}

\newcommand{\peasoup}{\textsc{Peasoup}}

\title[Reprocessing of the HTRU-S LowLat pulsar survey]{The High Time Resolution Universe Pulsar Survey – XIX. A coherent GPU accelerated reprocessing and the discovery of 71 pulsars in the Southern Galactic plane}

\author[Sengar et al.]{
R. Sengar$^{1,2,3,4,5}$\thanks{E-mail: sengar@uwm.edu},
M. Bailes$^{1,2}$,
V. Balakrishnan$^{6}$,
E. D. Barr$^{6}$,
N. D. R. Bhat$^{7}$,
M. Burgay$^{8}$,
\newauthor
M. C. i Bernadich$^{6}$,
A. D. Cameron$^{1,2}$,
D. J. Champion$^{6}$,
W. Chen$^{6}$,
C. M. L. Flynn$^{1,2}$,
A. Jameson$^{1,2}$,
\newauthor
S. Johnston$^{9}$,
M. J. Keith$^{10}$,
M. Kramer$^{6,10}$,
V. Morello$^{10}$,
C. Ng$^{11}$,
A. Possenti$^{8,12}$,
S. Stevenson$^{1,2}$,
\newauthor
R. M. Shannon$^{1,2}$,
W. van Straten$^{13}$,
J. Wongphechauxsorn$^{6}$
\\
$^{1}$ Centre for Astrophysics and Supercomputing, Swinburne University of Technology, Mail H39, PO Box 218, VIC 3122, Australia.\\
$^{2}$ARC Center of Excellence for Gravitational Wave Discovery (OzGrav), Swinburne University of Technology, Mail H11, PO Box 218, VIC 3122.\\
$^{3}$Center for Gravitation, Cosmology, and Astrophysics, Department of Physics, University of Wisconsin-Milwaukee, P.O. Box 413, Milwaukee, WI 53201, USA. \\
$^{4}$Max Planck Institute for Gravitational Physics (Albert Einstein Institute), D-30167 Hannover, Germany\\
$^{5}$Leibniz Universität Hannover, D-30167 Hannover, Germany\\
$^{6}$Max-Planck Institut f\"ur Radioastronomie, Auf dem H\"ugel 69, D-53121 Bonn, Germany.\\
$^{7}$International Centre for Radio Astronomy Research, Curtin University, Bentley, WA 6102, Australia.\\
$^{8}$ INAF - Osservatorio Astronomico di Cagliari, Via della Scienza 5, I-09047 Selargius (CA), Italy.\\
$^{9}$CSIRO Astronomy $\&$ Space Science, Australia Telescope National Facility, P.O. Box 76, Epping, NSW 1710, Australia.\\
$^{10}$Jodrell Bank Center for Astrophysics, University of Manchester, Alan Turing Building, Oxford Road, Manchester M13 9PL, United Kingdom.\\
$^{11}$Dunlap Institute for Astronomy \& Astrophysics, University of Toronto, 50 St.~George Street, Toronto, ON M5S 3H4, Canada.\\
$^{12}$Universit´a di Cagliari, Dept of Physics, S.P. Monserrato-Sestu Km 0,700 - 09042 Monserrato, Italy.\\
$^{13}$Institute for Radio Astronomy $\&$ Space Research, Auckland University of Technology, Private Bag 92006, Auckland 1142, New Zealand.\\
}

\date{Accepted XXX. Received YYY; in original form ZZZ}

\pubyear{2023}

\begin{document}

\label{firstpage}
\pagerange{\pageref{firstpage}--\pageref{lastpage}}\maketitle

\begin{abstract}

We have conducted a GPU accelerated reprocessing of $\sim 87\%$ of the archival data from the High Time Resolution Universe South Low Latitude (HTRU-S LowLat) pulsar survey by implementing a pulsar search pipeline that was previously used to reprocess the Parkes Multibeam pulsar survey (PMPS). We coherently searched the full 72-min observations of the survey with an acceleration search range up to $|50|\, \rm m\,s^{-2}$, which is most sensitive to binary pulsars experiencing nearly constant acceleration during 72 minutes of their orbital period. Here we report the discovery of 71 pulsars, including 6 millisecond pulsars (MSPs) of which five are in binary systems, and seven pulsars with very high dispersion measures (DM $>800 \, \rm pc \, cm^{-3}$). These pulsar discoveries largely arose by folding candidates to a much lower spectral signal-to-noise ratio than previous surveys, and exploiting the coherence of folding over the incoherent summing of the Fourier components to discover new pulsars as well as candidate classification techniques. We show that these pulsars could be fainter and on average more distant as compared to both the previously reported 100 HTRU-S LowLat pulsars and background pulsar population in the survey region. We have assessed the effectiveness of our search method and the overall pulsar yield of the survey. We show that through this reprocessing we have achieved the expected survey goals including the predicted number of pulsars in the survey region and discuss the major causes as to why these pulsars were missed in previous processings of the survey.


\end{abstract}




\section{Introduction}
\label{sec:intro}

The quest for new radio pulsars stands as a major and ongoing objective in the field of pulsar astronomy. Since their discovery in 1967 \citep{hbp68}, the tally of pulsars has climbed steadily and now reached 3389\footnote{\url{https://www.atnf.csiro.au/research/pulsar/psrcat/}} \citep[PSRCAT;][]{psrcat05a}. Pulsars have emerged as celestial beacons of immense scientific value as they have dramatically advanced our understanding in many areas of physics and astrophysics \citep{stairs_04}, such as in testing General Relativity and other alternative theories of gravity in the strong field regime using binary systems \citep[e.g.,][]{kramer_21}, providing evidence of a gravitational wave background \citep[e.g.,][]{reardon_23, nanograv_21a}, constraining the equation-of-state (EoS) using pulsar masses \citep[e.g.,][]{2020NatAs...4...72C}, testing theories of binary stellar evolution, and interactions in close binary systems \citep{tauris17}. Although millisecond pulsars (MSPs) play a crucial role in achieving the aforementioned goals, the majority of known pulsars, around 82\%, are normal pulsars with spin periods greater than about 30 milliseconds. These pulsars also provide insights into the enigmatic origin of pulsar emission which is still a topic of debate among researchers for decades \citep[e.g.,][]{2020PhRvL.124x5101P}. They also provide a window for mapping the Milky Way's pulsar population as well as extremes of the pulsar population, spanning from young to old, high to low magnetic fields, and slow to fast spins, offering valuable insights into the mechanisms behind radio pulse generation \citep[e.g.,][]{2018ApJ...866...54T}. The study of nulling and intermittent pulsars is essential for understanding pulsar emission mechanisms \citep[e.g.,][]{2011yera.confE..47Y, 2019JApA...40...42K}. Beyond their individual properties, pulsar distribution in the Galaxy is also crucial for understanding their population, beaming fraction, birth rate, local star-formation and supernova rates \citep[e.g.,][]{1994ApJ...437..781F}. Analyzing pulsar scale heights also refines formation models and velocities \citep{2002ApJ...568..289A}. Normal pulsars, abundant and observable at varying distances, offer crucial data for Milky Way cartography, aiding in understanding the Galaxy's structure, matter distribution, and stellar dynamics \citep[e.g.,][]{2022A&A...667A..82D}. Normal pulsar Rotation Measures (RMs) are vital for studying the Galactic magnetic field's large-scale structure \citep{han_99}. These pulsars also serve as probes for studying pulse scattering and enhancing Galactic electron density modelling \citep[e.g.,][]{ne2001, ymw16, ocker_20}.

Given the multitude of applications of both MSPs and normal pulsars, searching for them has become one of the key science goals for the future Square Kilometer Array \citep[SKA;][]{levin_18}. While the new pulsar population continues to grow by conducting large scale pulsar surveys with state-of-the-art radio telescopes such as the 500-m FAST in China  \citep{fast_gpps_21} and the 64-dish MeerKAT  radio telescope in South Africa \citep{padmanabh_23}, reprocessing of existing archival data sets has also been crucial in increasing pulsar population. A notable example of this practice is the analysis of the Parkes Multibeam Pulsar Survey (PMPS). Over the past two decades, with improvements in search techniques, candidate sifting, folding, and classification methods, the yield of the PMPS has increased by $\sim 50\%$ \citep{faulkner04,eatough13a,knispel13,mickaliger012} when compared to pulsar yield through its initial processing. Furthermore, advances in computing power, especially the utilization of graphical processing units (GPUs) in pulsar searches, have enabled a more thorough exploration of previously untapped search spaces at significantly faster rates \citep[e.g.,][]{morello19,sengar_23}. These reprocessing efforts have not only increased the pulsar population by discovering many interesting pulsars, but also resulted in a new class of pulsar called rotating radio transients \citep{mclaughlin06, keith011}. Another remarkable example of reprocessing pulsar survey is the discovery of the first Fast Radio Burst \citep[FRB;][]{lorimer_07} which was made through reprocessing archival data from a 1.4-GHz survey of the Magellanic Clouds using the multibeam (MB) receiver of the 64-m Parkes Radio Telescope \citep{manchester_06}.

In this paper, we present new results from a GPU-accelerated reprocessing of $87.3 \%$ of the HTRU-S LowLat survey
mainly motivated to search for accelerated (i.e. binary) pulsars. In Section \ref{sec:htru_survey_intro}, we give a brief introduction to the HTRU-S LowLat survey \citep{keith10}, touching on previous searches for relativistic binary pulsars and the
abundance of computational resources now available that provided
motivation for its reprocessing. Section \ref{sec:methods} provides an overview of data analysis methods applied to reprocess the 72-m integration length observations of the survey. Candidate confirmation and results, including new pulsar discoveries, are summarized in Section \ref{sec:discoveries}. Notable pulsar findings are further detailed in Section \ref{sec:notable discoveries}. The redetections of new pulsars in the PMPS survey are briefly described in Section \ref{sec:pmps_redetection}. In Section \ref{sec:statistical analysis}, we present a statistical comparison of the new pulsars against previous HTRU-S LowLat pulsars. Survey and pipeline efficiency are addressed in Section \ref{sec:pipeline_efficiency}. In Section \ref{sec:why_so_missed}, we give some insights into potential reasons why pulsars reported in this work elude their detection in the previous two processings of the survey. Finally, discussions and conclusions can be found in Section \ref{sec:discussion_and_conclusion}.

\section{HTRU-S LowLat pulsar survey and its reprocessing}
\label{sec:htru_survey_intro}

Over a decade ago three major pulsar surveys of the southern sky as part of the High Time Resolution Universe (HTRU) project \citep{keith10} were conducted using the 13-beam 20-cm multibeam (MB) receiver \citep{multibeam96} of the 64-m Parkes radio telescope (also knows as ``Murriyang'') between 2008--2013. The high-latitude ($\delta < +10^{\circ}$) part of the survey (HiLat) unveiled the cosmological population of fast radio bursts \citep[FRBs;][]{champion2016MNRAS}, the mid-latitude ($|b|<15^\circ, -120^{\circ}<l<30^{\circ}$) part of the survey (MedLat) found a rich pool of millisecond pulsars (MSPs) and a bright magnetar \citep{levin10}, whilst the low-latitude ($|b| <3.5^\circ$) part of the survey (HTRU-S LowLat) was tailored to detect relativistic binary pulsars and probe the low-luminosity end of the pulsar population. It is worth noting that while the PMPS had previously observed the same sky region using the MB receiver, the HTRU-S LowLat surpassed it in technological prowess.
This was achieved through a tenfold increase in frequency channelization, where the number of frequency channels increased from 96 in the PMPS to 1024 in the HTRU-S LowLat. While the total observing bandwidth was not significantly different i.e., from 288 MHz for PMPS to 340\footnote{the total HTRU-S LowLat bandwidth was 400 MHz with a channel bandwidth of 0.390 MHz.However, 150 channels were consistently affected by RFI, making the effective bandwidth to 340 MHz.} MHz for HTRU-S LowLat. Additionally, a finer time resolution was adopted, featuring a sampling rate of 64 $\mu \rm s$ compared to the PMPS's 250 $\mu \rm s$. The integration length was also extended, doubling from 36-min in the PMPS to a comprehensive 72-min in HTRU-S LowLat. These sensitivity enhancements made HTRU-S LowLat as technological successor to the PMPS, making it more sensitive to detect MSPs, highly scattered as well as fainter pulsars which potentially were missed or not detected in the PMPS.

The first search for the relativistic binary pulsars in 50$\%$ data of the HTRU-S LowLat survey was conducted by \citet{cherry15} with a CPU-based, partially coherent segmented acceleration search pipeline and discovered 60 pulsars. This pipeline used a ``time-domain resampling" acceleration search technique, which is sensitive when the length of the observation is 10\% of the orbital period of a binary system i.e. $t_{\rm int}/P_{\rm orb} \leq 0.1$ where, $t_{\rm int}$ is the integration time and $P_{\rm orb}$ is the orbital period of a pulsar in a binary system \citep{ransom_03}. Therefore, to detect highly accelerated binary pulsars, each 72-minute observation of the HTRU-S LowLat survey was divided into full-length, half-length, quarter-length and eighth-length segments, and an acceleration search (up-to a maximum acceleration of 1200\,$\rm m \, s^{-2}$) was performed separately on 15 segments of each observation. This segmentation provides sensitivity to binary pulsars with successively shorter orbital periods. Without employing this approach, potential highly accelerated pulsars might have remained undetected due to deleterious effects of Doppler smearing caused by the binary motion in large integration observations. Later, \citet{cameron2020high} processed the $44 \%$ of the data using the segmented acceleration search pipeline and discovered 40 pulsars. The first-pass processing of $94 \%$ of the HTRU-S LowLat data thus resulted in the discovery of 100 pulsars with 11 pulsars in binary orbits including one of the most accelerated binary pulsars, PSR J1757$-$1854 \citep{cameron_18}.

While the segmented search approach enhanced the sensitivity for highly accelerated pulsars experiencing noticeable jerk during 72-min of their orbit, there is still room for further refinement in search strategies. Notably, the segmentation search led to a reduction in sensitivity for the full-length observations, scaled by a factor of $\sqrt{n}$, where $n$ represents the number of segments (assuming an equal division). Although the full length observations were also coherently searched, they were only searched to mild accelerations within the range of $a=\pm  \,1 \rm m \, s^{-2}$. This limitation rendered the detection of faint, accelerated pulsars challenging. Additionally, prior to processing, all original observations in the filterbank format underwent down-sampling by a minimum factor of two, potentially diminishing sensitivity towards MSPs. Computational limitations at the time of the first search limited the phase space that could be explored.

With the development of GPU-based search codes in recent years, particularly on large-scale GPU clusters, the capacity to swiftly reprocess large amounts of pulsar survey data has become a reality. This technological leap has not only bolstered our ability to process data efficiently, but has also significantly broadened the parameter space available for exploration, resulting in an increase in the likelihood of discovering more pulsars \citep[e.g.,][]{morello19, sengar_23}. Therefore, with the expanded capacity of data processing, apart from the work presented in this paper, two parallel reprocessings of the survey were conducted using different search algorithms. The first involved Keplerian-parameter searches for PSR-BH binaries in circular orbits, utilizing a template bank algorithm. This technique led to the discovery of 21 new among which only one pulsar was identified as MSP in a binary system, while the others are isolated pulsars. The second method employed an acceleration search using the Fast Folding Algorithm \citep[FFA;][]{staelin_69}, but this analysis was restricted to process observations only around the Galactic Center. This processing resulted in the discovery of a new isolated pulsar, PSR J1746--2829, which has a large dispersion measure (DM) of 1309 $\rm pc \, cm^{-3}$ and is located just $0.5^{\circ}$ from the Galactic Center \citep{jompoj_23}. Note that all these pulsars were also redetected later in our reprocessing of the survey.

In this paper we will provide a comprehensive account of our reprocessing endeavours, leveraging GPU accelerators on the OzStar supercomputer at Swinburne University of Technology as we search for more pulsars from the HTRU-S LowLat survey.

\section{Methods and data analysis}
\label{sec:methods}

We developed a pipeline utilizing the GPU-based pulsar search code \texttt{PEASOUP} which was previously used to reprocess the PMPS survey \citep[see][]{sengar_23}. Since both the PMPS and HTRU-S LowLat surveys used the same multi-beam (MB) receiver, only minimal adjustments were required for the pipeline, primarily involving modifications to the parameter search space. The use of \texttt{PEASOUP} significantly optimized the most time-consuming tasks in CPU-based pulsar search pipelines such as the time-domain resampling and Fast Fourier search operations. In this study, \texttt{PEASOUP} was deployed on NVIDIA P100 12GB PCIe GPUs on the OzSTAR supercomputer and achieved an average wall-clock time of 22 ms per trial acceleration for HTRU-S LowLat observations. In comparison, CPU-based codes, such as the \texttt{seek} suite within \texttt{SIGPROC}, require approximately 19 seconds for similar operations. This represents an 850-fold increase in processing speed which is a substantial enhancement in computational efficiency using GPUs.

\subsection{RFI excision}
\label{subsec:RFI_excision}

The removal of radio frequency interference (RFI) was accomplished through a three-step process. Firstly, 155 out of the 1024 frequency channels were identified as contaminated by narrow band RFI and were always masked during dedispersion. Subsequently, two multibeam techniques were employed for RFI excision. In the time domain, an ``eigenvector decomposition'' method \citep{kocz10} was used, which involves producing a time series at $\rm DM=0\, pc \, cm^{-3}$ from each beam and forming a matrix for each time sample by cross-correlating these time series. The matrix was then decomposed into eigenvectors and eigenvalues, with the number of eigenvalues indicating the number of beams in which the signal was present. By applying a threshold, bad time samples contaminated by RFI could be identified and used as a mask in the time domain. In the Fourier domain, a periodic RFI mask was created using the ``multibeam coincidence technique" which involves calculating the power spectrum of 13 beams with $\rm DM=0\, pc \, cm^{-3}$  and applying a cutoff of 4$\sigma$ to remove any Fourier frequency present in four or more beams. The resulting list of periodic frequencies was called a periodic birdie list (`birdies' is a colloquial term for a form of RFI with a relatively pure tone or high Q value).

\subsection{Search parameters}
\label{subsec:search_params}

As outlined in Section \ref{sec:htru_survey_intro}, the previous two processings of the HTRU-S LowLat survey faced computational limitations in analyzing full-length data with native time resolution, especially in dealing with significant accelerations due to reliance on the CPU-based pipeline. However, the adoption of the GPU-accelerated pulsar search code \texttt{PEASOUP}\footnote{\url{https://github.com/ewanbarr/peasoup}} effectively overcame these constraints. This enabled us to explore wider ranges of acceleration using the survey's full-resolution dataset.

To perform coherent reprocessing, we employed the same GPU-based pipeline as discussed in \citet{sengar_23}, with the search parameters outlined in Table \ref{table:search_params}. Given that the Galactic centre is the densest region of free electrons in the Milky Way, we considered a dispersion measure (DM) search range extending up to a maximum DM of 2000 $\rm pc \, cm^{-3}$ with $\sim 2800$ trial DMs. This maximum DM is half of the highest line-of-sight predicted DM towards the Galactic center according to the NE2001 \citep{ne2001} and YMW16 \citep{ymw16} electron density models. However, this is consistent with the highest known DM pulsar SGR J1745−2900 which has a DM of 1778 $\rm pc \, cm^{-3}$. Therefore, we believe majority of the pulsars present detectable in the HTRU survey should have DM within our searched DM range.

\begin{table}
\caption{Search parameters used for re-processing the HTRU-S LowLat survey with \peasoup{}.} 
\centering 
\begin{tabular}{ l | l } 
\hline 
\hline
parameter & Value \\
[0.5ex] 
\hline  
$\rm DM_{\rm max} \: \rm (pc \: cm^{-3})$ & 2000 \\
$\rm N_{\rm DM trials}$ & 2778 \\
$\rm DM_{\rm tol}$ & 1.1 \\
$\rm W_{\rm int} \: (\mu \rm s)$ & 64 \\
$|\rm a_{max}| \: (\rm m \: \rm s^{-2})$ & 50\\
$\delta  a \: (\rm m \: \rm s^{-2}) $ & 0.05 \\
$N_{\rm acctrials}$ &  1884  \\
$N_{\rm harmonics}$ & 32 \\
$\rm Total \, Beams $ & 16153 \\
$\rm \, Beams \: \rm Processed$ & 14103 \\
$\rm Processing \: \rm time \, (GPU \, hr)$ & $\sim$ 430,000  \\
\hline 
\end{tabular}
\label{table:search_params}
\end{table}

At the onset of the processing, the version of \texttt{PEASOUP} we used loads the observation file (in filterbank format) into RAM and subsequently generates dedispersed time series for the specified number of DM trials using the \texttt{dedisp}\footnote{\url{https://github.com/ajameson/dedisp}} software package \citep{barsdel12}. Additionally, these time series are also stored in RAM prior to conducting periodicity and acceleration searches. However, loading all $\sim 2800$ time series would necessitate approximately 170 GB of RAM. Given that OzSTAR provided 192 GB of RAM per node at the time, allocating such a large portion for a single job would impede other CPU tasks. Therefore, we processed each observation into subsets of DM ranges, as shown in Table \ref{table:trials}.

\begin{table}
\caption{Number of DM trials used in each DM range in the reprocessing of the HTRU-S LowLat survey.}
\vspace{0.5cm}
    \centering 
    \begin{tabular}{ c | c }
    \hline  
    \hline
    DM range ($\rm pc \, cm^{-3}$) & no. of trial DMs\\
    [0.5ex] 
    \hline  
    0-50 & 357 \\
    50-100 & 270  \\
    100-200 & 346 \\
    200-500 & 503  \\
    500-800 & 265 \\
    800-1200 & 229 \\
    1200-2000  & 289  \\
    \hline 
    \end{tabular}
    \label{table:trials}
\end{table}

To search for binary pulsars, \texttt{PEASOUP} uses a time-domain resampling algorithm  \citep{middleditch_84, johnston92a}. This method achieves its highest sensitivity when the observation duration does not exceed a one-tenth of the pulsar's orbital period. Consequently, for the 72-minute HTRU-S LowLat observation, the sensitivity to linear acceleration is most pronounced for binary pulsars with orbital periods exceeding 12 hours, subject to a maximum constant acceleration of up to $|50| \, \rm m \, s^{-2}$, resulting in 1884 trials acceleration for each trial DM. Assuming circular orbits, this acceleration range is consistent with line-of-sight acceleration pulsar and 5 $M_{\odot}$ black-hole binary system and covers the line-of-sight acceleration observed in the majority of known binary pulsars. 

\subsection{Candidate selection and folding}
\label{subsec:sort_and_fold}

The candidate sifting and folding were also conducted using the methodologies explained in \citet{sengar_23}. The utilization of higher resolution data in modern pulsar surveys has resulted in a sharp increase in the volume of candidates generated during the search process. This increase poses a significant challenge in determining which candidates should be prioritized for folding, as folding tens of thousands of candidates per observation is infeasible. Traditionally, pulsar searches in the spectral domain employ a false alarm threshold that scales with the number of trials conducted during processing. Candidates falling below this threshold in the spectral domain are often considered to be products of random noise coincidences and hence rejected. However, \citet{sengar_23} demonstrated that adhering to a fixed false alarm threshold could lead to the overlooking of faint narrow-duty cycle pulsars when sifting through a substantial pool of candidates. This is because narrow periodic signals often exhibit power distributed across various harmonics in the Fourier domain. When these harmonics are incoherently summed to detect a peak in the power spectrum of the Fast Fourier Transformation (FFT), the resulting spectral Signal-to-Noise Ratio ($\rm S/N_{ FFT}$) is not optimal. In fact, for narrow duty cycle pulsars ($\delta<3.0 \%$), the folded Signal-to-Noise Ratio $\rm S/N_{ fold}$ can be 1.5 to 2.0 times higher compared to the $\rm S/N_{ FFT}$ \citep[see section 2.5.2 in][]{sengar_23} and for MSPs it remains more or less similar with $\rm S/N_{fold}$ being only $\sim$1.2 times higher. Considering this, we followed the candidate sifting technique proposed by \citet{sengar_23} and opted not to restrict candidate selection based on a fixed threshold. Instead, we employed varied cutoffs determined by parameters such as spin period, $\rm S/N_{ FFT}$, and the number of harmonics (nh), thus ensuring a comprehensive assessment of candidates, even those originating from the FFT noise floor.

Using these candidate filtering criteria, the initial pool of candidates was reduced by factor of about 10-12 per beam. Subsequently, a folding pipeline similar to the one employed by \citet{sengar_22} in reprocessing the PMPS was utilized for post processing. This pipeline leveraged the \texttt{dspsr}\footnote{\url{https://sourceforge.net/projects/dspsr/}} tool to fold candidates based on their designated period, dispersion measure (DM), and acceleration. We removed the known bad frequency channels from the data and used the \texttt{clfd}\footnote{\url{https://github.com/v-morello/clfd}} rfi cleaning software. The folding properties of the candidates such as $\rm S/N_{ fold}$, pulse width, optimized spin period, acceleration and DM as well as the diagnostic plots were obtained using \texttt{psrchive}'s tool, \texttt{pdmp} \citep{hotan_04}. Given that visually inspecting all folded candidates is a time-intensive and potentially error-prone task, we adopted the candidate discernment criteria outlined in section 3.1 of \citet{sengar_23}. This approach led to a substantial reduction in the candidate pool, enabling efficient one-person visual inspection within a very small fraction of a PhD candidacy.

\section{Candidate confirmation and New Discoveries}
\label{sec:discoveries}

We initially identified 98 promising pulsar candidates during the candidate inspection phase. Among these, 78 were designated as ``class A" candidates, meeting criteria such as visible and consistent signal in both frequency and time domain with folded $\rm S/N_{fold}$ above 8.8 for normal pulsar candidates and 12 for broad MSP candidates. The remaining 20 candidates fell into``class B", characterized by tentative visibility and $\rm S/N_{fold}$ below 8.8 or having broad pulse profiles with $\rm S/N_{fold}$ less than 12.

The confirmation and follow-up observations were conducted using the Parkes radio telescope. Initially, eight pulsars were confirmed using the MB receiver. However, due to its decommissioning in October 2020, the newly installed ultra-wide-band (UWL) receiver \citep{uwl_parkes}, spanning 704-4032 MHz in conjunction with the Medusa backend was used for confirming the remaining pulsar candidates. The confirmation process involved two distinct methods, similar to those utilized in recent reprocessing of the PMPS. The first method, the direct folding approach, employed a UWL frequency sub band (1200-1600 MHz) from observations and folded them using the detected period and DM of the candidates. This method successfully confirmed all normal pulsars. However, for MSP and accelerated pulsar candidates, an acceleration search was conducted to determine the optimal folding parameters. Additionally, the analysis of higher frequency bands provided by the UWL receiver proved instrumental in confirming several faint pulsars that might have otherwise been overlooked due to RFI in the 1200-1600 MHz frequency band, or would have necessitated additional confirmation attempts.

None of the class B candidates were confirmed as genuine pulsars. In contrast, out of the 78 class A candidates, 71 were confirmed as pulsars, reflecting 91\% confirmation success rate for class A candidates and 72\% overall success rate. This discrepancy in confirmation rates may be attributed to two potential factors. Firstly, the large number of trial accelerations, coupled with the DM trials, led to false-positive candidates with comparatively higher S/N levels, closely resembling real pulsar characteristics in their candidate plots. Discerning these false positives without direct observation posed a challenge. As for the seven unconfirmed class A slow pulsar candidates, they may potentially belong to the class of intermittent pulsars, necessitating multiple observations for definitive confirmation.

These new pulsars were overlooked in prior analyses of the HTRU-S LowLat survey, which we attribute to various factors detailed in Section \ref{sec:why_so_missed}. The basic properties of these newly confirmed pulsars, including spin period, dispersion measure (DM), Galactic coordinates from discovery observation, harmonic sums at which they were detected in the FFT power spectrum,  $\rm S/N_{FFT}$, and $\rm S/N_{fold}$ obtained from our pipeline, can be found in Table \ref{table:new_pulsars_table}. For each discovery pulse profile of the pulsar, we fit a Gaussian model using \texttt{lmfit} software package to each profile and reported the pulse width at 50\% ($\rm W_{50}$) of the peak pulse (see Table \ref{table:pulsar_distance_table}). Furthermore, Figure \ref{fig:pulse profiles} showcases a composite view of integrated pulse profiles from the discovery observations.

\begin{table*}
\caption{{Discovery parameters of 71 confirmed new pulsars from $\sim$ 87.3 $\%$ reprocessing of the HTRU-S LowLat survey. The right ascension (RA) and declination (DEC) of these pulsars are the coordinates of the beam centre in which the pulsar was found. The positional error in declination are estimated from the beam size of the MB receiver. The barycentric (BC) spin period ($P$) in ms and DM values are listed from the discovery observations. The number of harmonics (nh) reported here correspond to the harmonic sums in which the candidate was detected by our pipeline. The paramteres $\rm S/N_{\rm FFT}$, $\rm S/N_{\rm HTRU-S}$ correspond to the SNs of the new pulsars obtained in the FFT search and optimized SN after folding. $\rm S/N_{\rm PMPS}$ is the folded SN of pulsar redetection in the PMPS.}}

\label{table:new_pulsars_table}
\centering
\hspace*{-0.5cm}
\begin{tabular}{llllllllll}

\Xhline{1.5\arrayrulewidth}
PSR name & pointing/beam & RA & DEC  & $\rm P$ & $\rm DM$ & nh&$\rm S/N_{FFT}$ & $\rm S/N_{HTRU-S}$&$\rm S/N_{PMPS}$ \\
 & & (hh:mm:ss)& (dd:mm:ss)& $\rm (s)$ & $\rm (pc \, cm^{-3})$ & & & &\\ 

\Xhline{1.5\arrayrulewidth}

J1136$-$64 & 2012-07-23-00:10:48/10 & 11:36.4(5) & $-$64:40(7) &1.0236193(31)&309(10)&32&9.0&14.0& -    \\
J1306$-$60 & 2011-12-27-15:02:44/13 & 13:6.2(5) & $-$60:21(7) &1.783181(15)&283(28)&32&6.2&11.5& 8.2$^{c}$  \\
J1310$-$63 & 2012-04-02-09:33:32/10 & 13:10.3(5) & $-$63:19(7) &0.18566668(10)&625(2)&16&7.2&9.7& -    \\
J1325$-$6253$^{\dagger}$ & 2011-12-10-16:54:46/10 & 13:25.04.890 & $-$62:53:39.594 &0.0289706736(24)&303(0)&8&10.4&14.2& -    \\
J1333$-$61 & 2011-09-18-22:03:34/07 & 13:33.6(5) & $-$61:48(7) &1.5321347(69)&546(15)&32&12.0&17.3& 9.4$^{c}$   \\
J1348$-$62 & 2011-10-10-21:55:27/06 & 13:48.0(5) & $-$62:30(7) &0.6162574(56)&790(30)&2&7.7&11.0& 7.7$^{c}$   \\
J1406$-$59 & 2011-04-19-15:27:52/07 & 14:6.9(5) & $-$59:23(7) &1.2483120(46)&286(12)&32&9.3&12.8& 9.0$^{c*}$  \\
J1423$-$62 & 2011-04-23-09:21:43/05 & 14:23.6(5) & $-$63:45(7) &1.671473(10)&382(19)&16&5.2&10.0& -    \\
J1437$-$62 & 2011-12-05-20:04:47/06 & 14:37.10(5) & $-$62:47(7) &0.7779940(17)&323(7)&16&8.6&14.1& 10.2$^{c*}$ \\
J1445$-$63 & 2012-07-31-11:22:27/12 & 14:49.8(5) & $-$63:45(7) &0.003718668530(46)&79(0)&8&13.7&16.0& -    \\
J1518$-$60 & 2012-07-31-12:35:11/03 & 15:18.2(5) & $-$60:29(7) &0.51065557(77)&420(5)&8&20.2&15.0& -    \\
J1521$-$57 & 2013-02-02-16:02:33/12 & 15:21.8(5) & $-$57:47(7) &0.173681695(89)&270(2)&8&7.7&10.5& -    \\
J1549$-$5337$^{\rm a}$ & 2011-05-18-11:12:03/13 & 15:49.0(0.5) & $-$53:37.4(0.7)&0.003316709547(36)&184(0)&4&11.2&13.4& -    \\
J1548$-$55 & 2011-12-29-17:55:52/06 & 15:48.5(5) & $-$55:54(7) &0.5413823(18)&446(11)&8&5.7&9.0& -    \\
J1555$-$53 & 2011-05-17-11:18:22/03 & 15:55.0(5) & $-$53:00(7) &1.1708960(88)&785(25)&4&9.1&12.0& 10.5$^{c*}$ \\
J1559$-$55 & 2012-08-07-03:38:29/05 & 15:59.9(5) & $-$55:37(7) &1.3007370(50)&345(13)&32&7.0&10.0& 7.9$^{c}$  \\
J1600$-$49 & 2012-01-04-18:26:39/04 & 16:0.4(5) & $-$49:39(7) &0.28752610(41)&328(5)&8&7.5&10.6& -    \\
J1603$-$54 & 2011-12-07-21:23:21/04 & 16:3.4(5) & $-$54:05(7) &0.9607924(27)&472(9)&32&8.3&12.6& 11.1$^{c*}$ \\
J1605$-$52 & 2012-08-03-05:51:26/10 & 16:5.2(5) & $-$52:28(7) &2.193373(14)&461(21)&16&5.3&9.2& 8.4$^{c*}$  \\
J1614$-$52$^{\rm b}$ & 2012-11-26-21:55:56/07 & 16:14.7(5) & $-$52:20(7) &0.5099027(18)&673(12)&8&9.2&10.5& 10.1$^{c*}$    \\
J1628$-$46 & 2011-12-05-03:39:15/04 & 16:28.0(5) & $-$46:25(7) &0.44943888(77)&513(5)&8&5.1&9.8& -    \\
J1631$-$47 & 2011-12-30-19:34:06/05 & 16:31.10(5) & $-$47:10(7) &1.1033263(42)&756(13)&8&7.1&10.4& 9.4$^{c*}$  \\
J1632$-$49 & 2011-12-13-04:39:04/12 & 16:32.6(5) & $-$49:05(7) &0.4168363(16)&814(13)&4&12.3&12.5& -    \\
J1634$-$49 & 2011-04-23-20:03:59/02 & 16:34.6(5) & $-$49:58(7) &0.35671537(37)&465(3)&16&12.1&15.0& -    \\
J1635$-$46 & 2011-06-26-16:09:12/11 & 16:35.9(5) & $-$46:45(7) &1.4889028(65)&549(14)&32&10.7&14.9& 8.6$^{c}$  \\
J1638$-$47 & 2012-04-14-13:41:06/02 & 16:38.0(5) & $-$47:50(7) &0.4266682(24)&1374(19)&2&13.2&12.5& 12.5$^{c*}$ \\
J1639$-$46 & 2012-04-12-12:47:43/13 & 16:39.2(5) & $-$46:27(7) &0.5191365(10)&917(7)&4&6.9&9.1& -    \\
J1641$-$49 & 2011-10-13-06:08:35/09 & 16:41.8(5) & $-$49:34(7) &0.7951897(23)&578(9)&16&5.3&9.9& 7.6$^{c}$  \\
J1647$-$49 & 2011-10-10-05:13:04/06 & 16:47.3(5) & $-$49:10(7) &0.24752785(18)&515(2)&16&8.7&12.6& 9.2$^{c}$  \\
J1652$-$4237$^{\rm a}$ & 2012-12-30-02:45:46/07 & 16:52.0(0.5) & $-$42:37.5(0.7) &0.4965476(21)&943(14)&2&12.7&12.4& 15.7$^{c*}$ \\
J1651$-$46 & 2011-12-07-22:36:35/04 & 16:51.1(5) & $-$46:47(7) &0.5693516(19)&487(11)&4&10.8&13.4& -    \\
J1655$-$40 & 2011-04-20-20:48:56/03 & 16:55.2(5) & $-$40:10(7) &0.27668901(50)&439(6)&2&9.7&11.9& 10.5$^{c*}$ \\
J1700$-$39 & 2011-12-10-04:41:41/01 & 17:0.10(5) & $-$39:43(7) &3.746463(41)&509(36)&32&10.1&13.5& 7.9$^{c}$  \\
J1708$-$38 & 2012-03-30-21:22:02/09 & 17:8.9(5) & $-$38:04(7) &0.6698359(13)&526(6)&8&6.8&9.3& -    \\
J1710$-$3946$^{\rm a}$ & 2011-12-29-20:22:33/01 & 17:10.8(0.5) & $-$39:46.3(0.7) &0.9773371(88)&1198(29)&4&6.8&12.8& 8.6$^{c}$  \\
J1717$-$41 & 2011-10-13-07:22:02/04 & 17:17.4(5) & $-$41:19(7) &0.54623299(88)&356(5)&16&12.9&15.4& 9.2$^{c*}$  \\
J1719$-$36 & 2012-08-03-08:18:01/09 & 17:19.9(5) & $-$36:42(7) &0.7571511(16)&624(7)&16&10.1&11.8& 7.6$^{c}$  \\
J1723$-$40 & 2012-11-25-01:22:55/13 & 17:23.2(5) & $-$40:47(7) &1.982265(11)&347(19)&16&11.7&14.0& 11.1$^{c*}$    \\
J1735$-$28 & 2011-06-27-16:02:05/12 & 17:35.2(5) & $-$28:36(7) &0.42856014(65)&279(5)&16&7.4&10.0& -    \\
J1735$-$33 & 2012-09-30-05:41:45/11 & 17:35.10(5) & $-$33:37(7) &1.2738657(48)&171(12)&32&7.7&10.6& -    \\
J1738$-$33 & 2012-10-04-10:29:13/09 & 17:38.9(5) & $-$33:52(7) &0.35773440(37)&257(3)&8&6.6&9.7& -    \\
J1739$-$26 & 2011-06-30-16:47:36/07 & 17:39.4(5) & $-$26:23(7) &0.49014182(78)&354(5)&8&7.2&10.6& -    \\
J1755$-$22 & 2011-04-21-19:06:17/08 & 17:55.3(5) & $-$22:49(7) &0.57068731(94)&273(5)&32&6.57&8.9& -    \\
J1756$-$25 & 2011-04-24-21:02:42/11 & 17:56.2(5) & $-$25:18(7) &0.48788974(70)&190(5)&32&6.9&9.4& -    \\
J1757$-$26 & 2011-10-06-07:02:33/12 & 17:57.1(5) & $-$26:52(7) &0.3544587(26)&502(24)&1&13.9&13.0& 9.8$^{c*}$    \\
J1758$-$24 & 2011-04-24-21:02:42/03 & 17:58.3(5) & $-$24:27(7) &0.6331055(16)&652(9)&4&12.4&13.5& 8.3$^{c}$  \\
J1758$-$25 & 2013-04-04-17:59:06/13 & 17:58.9(5) & $-$25:27(7) &0.6055017(10)&415(6)&32&6.6&9.9& 7.6$^{c}$  \\
J1806$-$2133$^{\rm a}$ & 2013-01-07-04:18:23/08 & 18:06.2(0.5) & $-$21:33.9(0.7) &0.3285584(14)&989(14)&16&8.9&12.9& -    \\
1808$-$14 & 2013-04-06-16:18:54/03 & 18:8.2(5) & $-$14:57(7) &0.8364221(20)&307(8)&16&7.2&12.0& -    \\
J1808$-$19 & 2011-07-02-12:25:09/03 & 18:8.1(5) & $-$19:37(7) &0.10166610(10)&969(3)&1&14.2&16.5& -    \\
J1809$-$20 & 2013-04-02-16:32:16/06 & 18:9.8(5) & $-$20:36(7) &0.057256207(22)&528(1)&2&8.9&11.9& -    \\
J1812$-$12 & 2011-12-29-00:45:10/07 & 18:12.10(5) & $-$12:54(7) &1.4399673(63)&278(14)&32&5.1&9.7& 6.8$^{c}$  \\
J1813$-$14 & 2013-04-02-21:22:53/04 & 18:13.5(5) & $-$14:44(7) &1.0354074(31)&427(10)&16&9.0&13.4& 9.3$^{c}$  \\
J1814$-$1845$^{\rm a}$ & 2013-04-08-18:54:20/05 & 18:14.7(0.5) & $-$18:45.4(0.7)&1.0899895(35)&534(10)&8&36.1&33.4& 8.5$^{c}$  \\
J1817$-$19 & 2013-04-08-18:54:20/10 & 18:17.1(5) & $-$19:35(7) &1.2290849(44)&191(12)&32&9.0&12.8& -    \\
J1820$-$19 & 2012-11-27-01:35:49/04 & 18:20.6(5) & $-$19:46(7) &0.004490802565(90)&142(0)&4&15.8&18.8& -    \\
J1820$-$20 & 2012-12-10-01:36:29/06 & 18:20.8(5) & $-$20:55(7) &0.002699801845(64)&274(0)&4&17.0&18.2& -    \\
J1823$-$11 & 2011-12-31-00:27:33/06 & 18:23.8(5) & $-$11:48(7) &0.2861844(13)&725(15)&2&9.3&11.0& -    \\
J1830$-$09 & 2011-07-05-14:23:13/09 & 18:30.6(5) & $-$9:08(7)  &0.6954903(14)&309(7)&32&11.2&15.9& -    \\
J1830$-$14 & 2011-05-07-15:44:36/13 & 18:30.1(5) & $-$14:34(7) &0.39984826(94)&297(8)&8&6.4&9.2& 8.8$^{c}$  \\

\Xhline{1.5\arrayrulewidth}
\end{tabular}
\end{table*}

\begin{table*}
\caption*{continued}
~\\
\centering 
\resizebox{1.0\textwidth}{!}{%
\begin{tabular}{cccccccccc}
\hline

\Xhline{1.0\arrayrulewidth}
PSR name & pointing/beam & RA & DEC  & $\rm P$ & $\rm DM$ & nh&$\rm S/N_{PMPS}$ & $\rm S/N_{HTRU}$&$\rm S/N_{PMPS}$ \\
 & & (hh:mm:ss)& (dd:mm:ss)& $\rm (ms)$ & $\rm (pc \, cm^{-3})$ & & & &\\ 
\hline
\Xhline{1.0\arrayrulewidth}
J1833$-$05 & 2012-04-14-19:48:16/08 & 18:33.2(5) & $-$05:13(7) &0.7448830(19)&380(8)&16&6.6&11.0& -    \\
J1835$-$09 & 2013-04-01-18:01:40/11 & 18:34.6(5) & $-$9:15(7)  &0.7504618(50)&555(22)&16&10.5&12.5& 10.9$^{c*}$ \\
J1837$-$10 & 2011-07-14-15:08:44/13 & 18:37.3(5) & $-$10:13(7) &1.0166051(36)&493(12)&16&6.0&9.7& -    \\
J1840$-$09 & 2013-04-01-19:14:02/07 & 18:40.6(5) & $-$09:37(7) &0.003089100596(94)&169(0)&4&10.5&11.6& -    \\
J1841$-$05 & 2012-08-03-13:11:46/13 & 18:41.0(5) & $-$05:19(7) &1.0885105(35)&274(10)&32&6.0&10.4& -    \\
J1842$-$06 & 2012-08-05-09:30:53/07 & 18:42.8(5) & $-$06:21(7) &0.36077664(46)&539(4)&8&7.3&10.4& -    \\
J1844$-$02 & 2011-01-03-00:48:46/06 & 18:44.8(5) & $-$2:40(7)  &0.5815353(12)&336(7)&16&8.5&11.6& 8.4$^{c}$  \\
J1844$-$09 & 2013-04-01-20:26:44/01 & 18:44.2(5) & $-$9:10(7)  &0.6344454(11)&415(6)&16&8.5&12.2& 10.2$^{c*}$ \\
J1846$-$05 & 2012-04-13-20:25:11/06 & 18:46.2(5) & $-$5:29(7)  &1.4449846(61)&474(14)&32&10.1&14.5& 7.5$^{c}$  \\
J1847$-$05 & 2011-12-23-03:12:09/01 & 18:47.7(5) & $-$05:41(7) &2.618072(20)&170(25)&32&7.4&13.3& -    \\
J1854$-$05$^{\rm b*}$ & 2011-12-13-06:05:17/01 & 18:54.2(5) & $-$05:12(7)  &1.2799447(48)&275(12)&32&9.4&12.4& 11.4$^{c}$ \\
\Xhline{1.0\arrayrulewidth}
\hline
\label{}
\end{tabular}}

\raggedright{
\small{${^{\dagger}}$ A double neutron star system \citep{sengar_22}\\
\small{${^{\rm a}}$ Positions were obtained using MeerKAT's FBFUSE and APSUSE backends \citep{trapum_16}. The localization was performed using 480 MeerKAT beams tiled across the area of the MB receiver, and the coordinates of the beam where the pulsar was best detected with the highest SN are reported.}\\

\small{${^{\rm b}}$ Later independent discovery in MMGPS-S as PSR J1614$-$5218, \url{http://www.trapum.org/discoveries/}.}\\

\small{${^{\rm b*}}$ Independent discovery in the Commensal Radio Astronomy FAST Survey (CRAFTS), \url{http://groups.bao.ac.cn/ism/CRAFTS/202203/t20220310_683697.html}.}\\
\small{${^{\rm c}}$ Pulsars found using direct folding}.\\
\small{${^{\rm c*}}$ Pulsars found in both direct folding as well as in the FFT search}.\\}}

\end{table*}

\section{Notable discoveries}
\label{sec:notable discoveries}

Among the 71 new pulsars reported in this work, PSR J1325--6253 is a double neutron star system and has remarkably low orbital eccentricity of 0.064, especially given its 1.81 d long orbit \citep{sengar_22}. Of the other newly identified pulsars, five are MSPs, with four of them being part of binary systems. The remaining pulsars are categorized as normal, with the slowest one having a spin period of 3.7 seconds. Interestingly, four of these normal pulsars possess relatively fast spin periods (ranging from 50 ms to 200 ms), suggesting they may be part of a younger pulsar population. Additionally, one normal pulsar exhibits nulling, and seven pulsars have very high dispersion measures (DM), surpassing 800 $\rm pc  \, cm^{-3}$. The timing solutions for these pulsars will be addressed in upcoming publications. Therefore, below we provide a brief overview of these pulsars. Note that orbital parameters for binary MSPs were determined using \texttt{fitorbit}\footnote{\url{https://github.com/vivekvenkris/fitorbit}}.

\subsection{Millisecond pulsars}

\subsubsection{PSR J1445--63 }

PSR J1445$-$63 is 3.71 ms pulsar with a DM of $\rm 79.5\, pc \, cm^{-3} $. In our search, it was first identified as a candidate in one of the observations conducted on UTC 2012-07-31-11:22:27, with a mild acceleration indicating the possibility of it being in a binary system. However, several initial attempts to confirm the pulsar were unsuccessful. Subsequently, during the candidate inspection stage, PSR J1445$-$63 was again detected in two adjacent observations with a mild acceleration, which further indicated that it is a real pulsar in a binary system. A large positional offset between the three observations suggested that the pulsar was either very bright or that its position was recorded inaccurately in the survey metadata. Upon further investigation, erroneous data in the PSRXML header file was discovered, and the corresponding observation was rejected as a probable position of the pulsar. To determine the pulsar's accurate position, a gridding technique was employed between the two closest observations, and the pulsar was observed for an hour, and we detected it with a high $\rm S/N_{fold}$ of 35. \par

The subsequent observations of the pulsar revealed that it is indeed a binary pulsar in  a 14.09 d orbit with a companion with minimum and median masses of 0.23 $\rm M_{\odot}$ and 0.27 $\rm M_{\odot}$, respectively. Additionally, the pulsar exhibits two distinct pulses in its profile, with the interpulse pulse separated from the leading pulse by approximately 194 degrees.

\subsubsection{PSR J1840--09}

PSR J1840--09 is 3.089 ms pulsar with a DM of 169.9 $\rm  pc \, cm^{-3} $. It is present in a binary system and has an orbital period of 5.47 d with a companion with minimum and median masses of 0.15 $\rm M_{\odot}$ and 0.17 $\rm M_{\odot}$, respectively. The pulse profile of this pulsar also shows a strong post-cursor pulse which is separated by about 110 degrees from its leading pulse.

\subsubsection{PSR J1549--5337}

PSR J1549$-$5337 has a spin period of 3.31 ms and the widest orbit of 66 d among the five MSPs reported in this work. Its DM of 184 $\rm pc \, cm^{-3}$ suggests a distance of 3.9 kpc and 3.5 kpc based on the NE2001 \citep{ne2001} and YMW16 \citep{ymw16} electron density models, respectively. The minimum companion mass, assuming an orbital inclination angle of $i=90^{\circ}$ and pulsar mass of 1.4 $\rm M_{\odot}$, is 0.20 $\rm M_{\odot}$, while the median mass assuming $i=60^{\circ}$ is 0.23 $\rm M_{\odot}$. Despite multiple observations over 219 d, its relatively wide orbit still makes phase connection incomplete and high cadence observations will be required in the future. However, with precise position measurements obtained using the MeerKAT radio telescope \citep{trapum_16}, we found that PSR J1549$-$5337 has a narrow pulse profile with a Gaussian full-width half-maxima of the pulse, $W_{50}$, of 0.14$\pm$0.01 ms, corresponding to a duty cycle of $4.3\pm0.2\%$ whereas the median duty cycle of MSPs with $P<30 \, \rm ms$ is $\sim 9.3\%$. Continuous monitoring with sensitive telescopes, such as MeerKAT, can provide improved timing accuracy, making it a promising candidate to be included in pulsar timing arrays to detect gravitational waves unless its high DM makes precision timing difficult. 

\subsubsection{PSR J1820--20}

PSR J1820$−$20 has a spin period of 2.69 ms which is the fastest among new MSPs reported in this work. It has a DM of 275.1 $\rm pc \, cm^{-3}$, and based on the NE2001 and YMW16 electron density models, it is located at a distance of 5.7 kpc and 10.0 kpc, respectively. This pulsar is in a binary system with an orbital period of 16.3 d, and its companion has minimum and median masses of 0.23 $\rm M_{\odot}$ and 0.27 $\rm M_{\odot}$, respectively.

\subsubsection{PSR J1820$-$19}

PSR J1820$-$1946 is an isolated MSP with a spin period of 4.49 ms and DM of 143 $\rm pc \, cm^{-3}$. Again using the NE2001 and YMW16 electron density models, its distance is 3.1 kpc and 3.2 kpc, respectively. One of the noticeable features of PSR  J1820$-$1946 is that it is detectable across almost the entire UWL band (1000-4000 MHz). The pulsar is highly scattered from 1000 MHz to 1500 MHz, and at higher frequencies, the pulse profile is narrower as expected. The formation of isolated MSPs is not well understood and highly debated; however, there are a variety of different formation scenarios. Most plausible scenarios include the formation of massive isolated MSPs via the merging of a NS and a massive WD \citep[e.g.,][]{vanden_84,bailes_11} or the companion's ablation due to the pulsar wind \citep[e.g.,][]{fruchter_90}.

\subsection{PSR J1518--60: A nulling pulsar}

\begin{figure}
\centering
\includegraphics[width=0.4\textwidth]{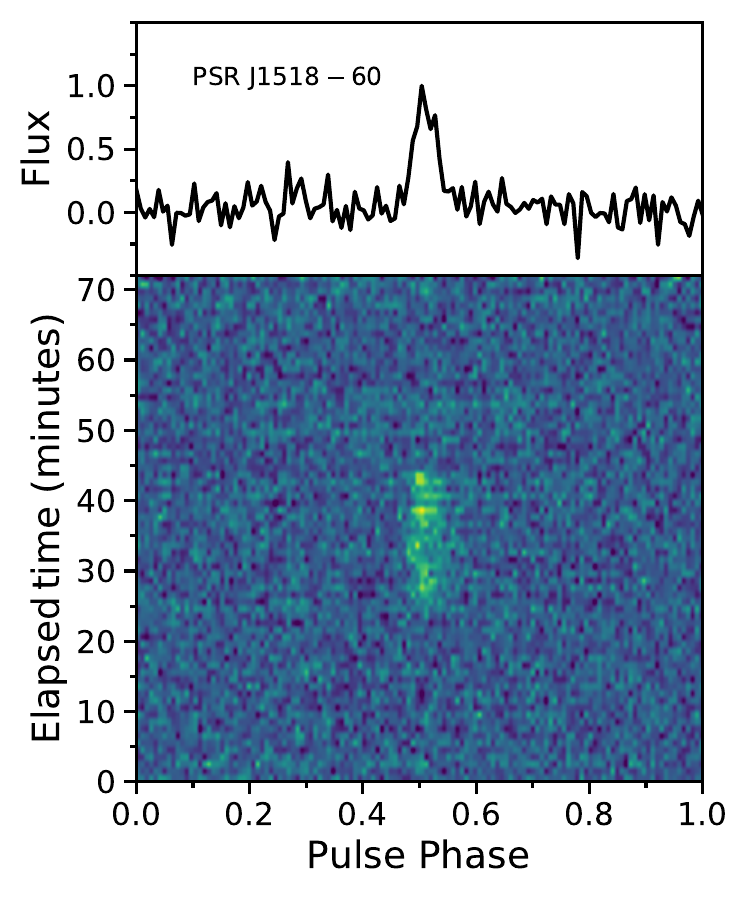}
\caption{Top panel shows the integrated profile of the nulling pulsar PSR J1518--60 and bottom plot shows the time-phase plot where the pulsar is visible for only $\sim$ 25 minutes.}
\label{fig:nulling_psr_plot}
\end{figure}

PSR J1518$-$60 exhibits evidence of nulling in its discovery observation and was detected with the high S/N of 15. It shows broadband emission across the entire 350 MHz band and visibility in only $\sim 25 \%$ of the observation (see Figure \ref{fig:nulling_psr_plot}). 
Apart from this, its relatively high DM of 419 $\rm pc \, cm^{-3}$ is constrained well away from 0 $\rm pc \, cm^{-3}$ with a DM error of only 5 $\rm pc \, cm^{-3}$, bolstering the case  that J1518$-$60 is indeed a real pulsar. There have been no further detections of it in 2.9 hr observations taken between MJDs 59216$-$59246 since its initial discovery in the survey, suggesting that it may have a large nulling fraction. Based on the method described in \cite{lyne17}, the fraction of time during which the pulsar is in the ``$\rm ON$'' state, i.e. nulling fraction (NF), can be given by $t_{\rm ON}/t_{\rm total}$ where $t_{\rm ON}$ and $t_{\rm total}$ correspond to the time during which the pulsar is visible in radio or in `ON' state and the total observation time, respectively. For J1518$-$60, the NF in this observation is $73(17) \%$. To further detect this system, observations with higher cadence or those extending over longer durations will be required.

\begin{figure*}
\vspace{-0.5cm}
\begin{center}
\includegraphics[width=1.8\columnwidth,angle=0]{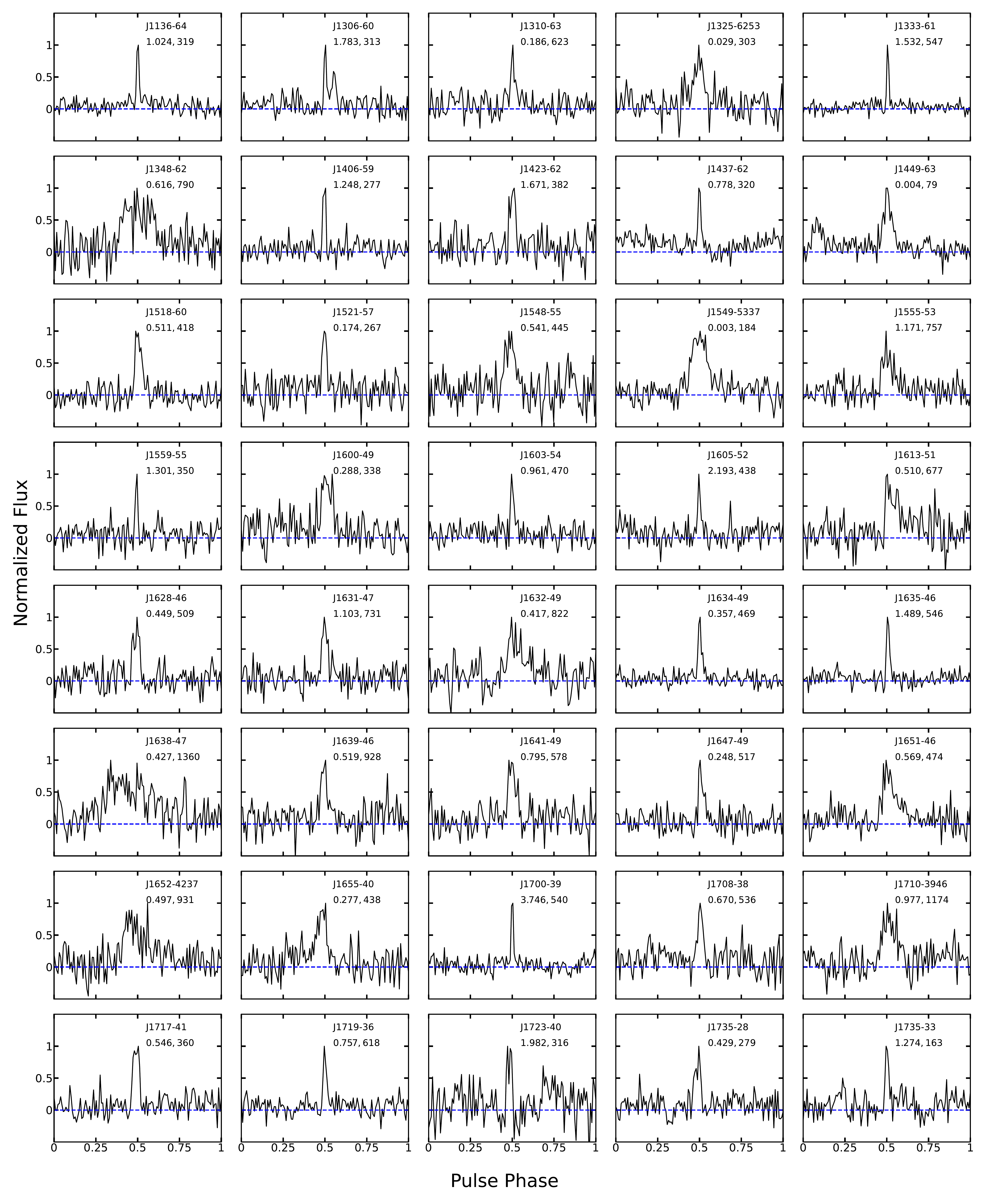}
\caption{
Integrated pulse profiles for  the newly discovered pulsars. Each profile contains 128 bins and has been rotated such that the it peaks at 0.5 phase. The period in seconds and the DM in $\rm pc \, cm^{-3}$ are provided for each pulsar.}
\label{fig:pulse profiles}
\end{center}
\end{figure*}

\begin{figure*}
\vspace{-0.5cm}
\begin{center}
\includegraphics[width=1.8\columnwidth,angle=0]{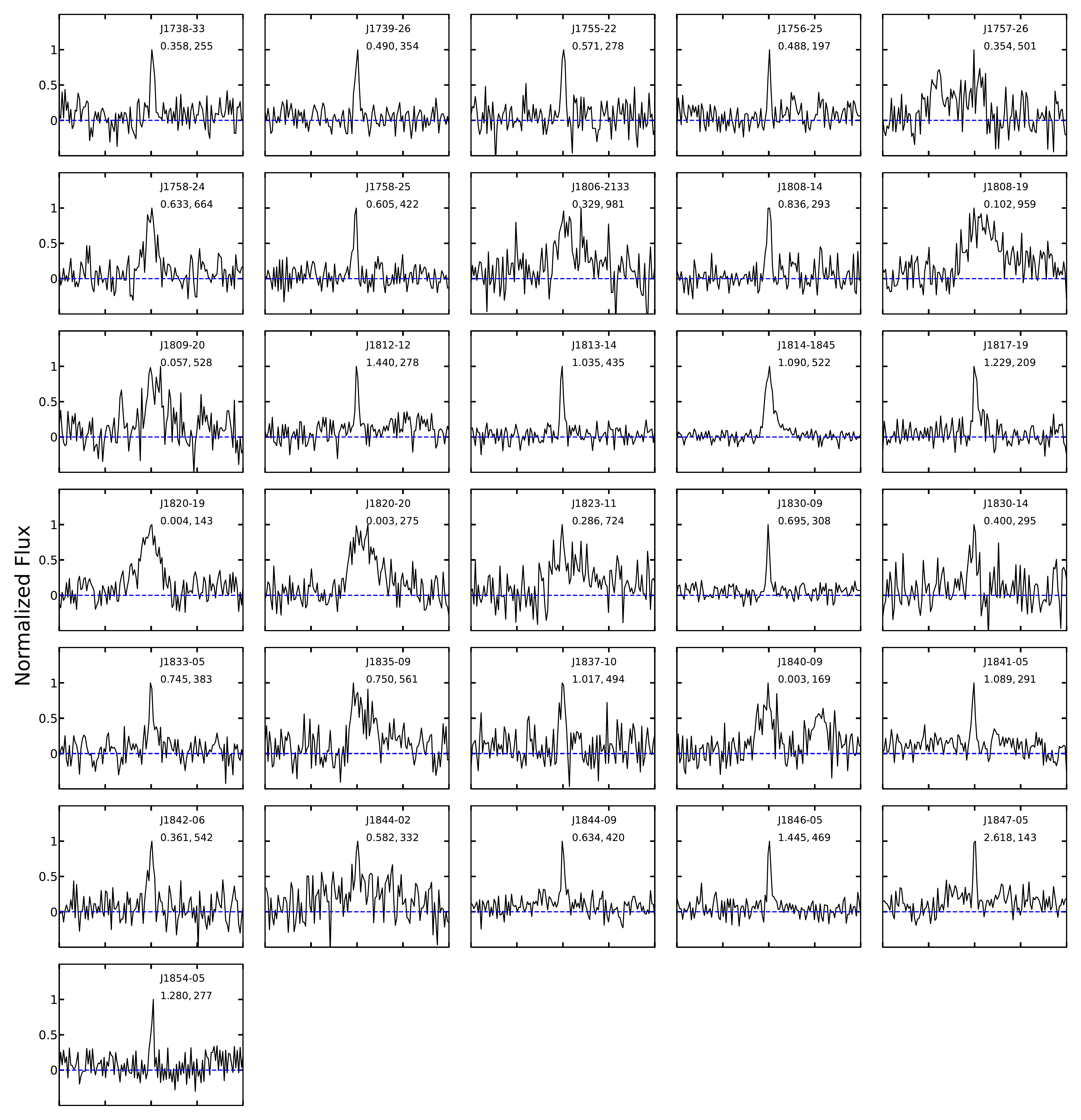}
\caption*{continued}
\end{center}
\end{figure*}

\subsection{A collection of high DM pulsars}

Apart from MSPs and a nulling pulsar, there is another class of pulsars which we term ``high DM'' pulsars. In the new pulsar sample, seven pulsars have $\rm DM>800 \,  pc \, cm^{-3}$ among which PSR J1638$-$47 and J1710$-$39 have a DM of 1374(19) and 1198(29) $\rm pc \, cm^{-3}$, respectively, and are among the top ten highest DM pulsars in the known pulsar population. These two pulsars are present in the PMPS data but due to low SNR and scattered pulse they remained undetected in previous multiple processings of the survey, however recently they were reported as an independent discovery by \citet{sengar_23}. The large DM of these pulsars is not surprising as these pulsars are located along the tangential direction of the Galactic spiral arms. PSR J1638$-$47 is also very close to the maximum DM predicted by YMW16 electron density model, but it is less than the maximum DM predicted by NE2001 electron density model. Similarly, PSR J1651--42 ($\rm DM \sim 931 \rm pc \, cm^{-3}$) exceeds the maximum DM prediction by YMW16, but it is less than DM predicted by NE2001. There are five more pulsars whose DM is within the 80\% of the maximum line-of-sight DM predicted by either of the NE2001 or YMW16 electron density models. Since none of the pulsar exceeds the maximum DM prediction by both models suggests that they are within the Milky Way.

\section{Redetection of new pulsars in the PMPS survey}
\label{sec:pmps_redetection}

As the entire portion of the Galactic plane surveyed in the HTRU-S LowLat survey overlaps with the region observed in the PMPS, a search for the closest PMPS observations was warranted to determine if any of the 71 new pulsar discoveries reported in this work are also detectable in the PMPS. Two separate searches were conducted for the PMPS observations within one beam-width (14.4$^{\circ}$) of the MB receiver from the original discovery position of the HTRU-S LowLat observations. First, we used the ``direct folding method'' in which the PMPS observations were folded with the original discovery period and DM of the pulsars, and on optimizing the periods and DMs of the pulsars, if the pulsar signal is visible above a $\rm S/N_{fold}$ of $\sim$ 7, then it was considered as a redetection. Of the 71 pulsars, 34 normal pulsars were redetected using the direct folding method, among which 28 were detected with $\rm S/N_{fold}$ $\geq 8$ \citep{pmps01}. The remaining 6 pulsars have low detection significance with $\rm S/N_{fold}$ between 7 and 8. Similarly, following the same methodology of direct folding, in the previous analysis of the HTRU-S LowLat survey conducted by \cite{cherry15} and \cite{cameron2020high}, 31 out of 100 pulsars were redetected above this significance level. Therefore, a total of at least 65 pulsars out of 171 are redetected in the PMPS survey using the direct folding method.

In the second approach, a blind FFT search was conducted on the PMPS observations for the 71 new HTRU-S LowLat pulsars. This search spans a dispersion measure (DM) range from 0 to 1500 $\rm pc \, cm^{-3}$. Out of the 34 pulsars detectable through the direct folding method (see Table \ref{table:new_pulsars_table}), approximately 45\% (15 pulsars) were detected, and  meet the criterion of having $\rm S/N_{fold}$ greater than or equal to 8.5. However, upon reviewing their candidate plots, ten pulsars stand out distinctly and are clearly recognizable. Theoretically, these pulsars should have been detected in previous reprocessings of the PMPS survey. However, for unknown reasons, they remained undetected. It is possible that due to the large number of candidates, they were overlooked during candidate sorting and inspection, or not picked up in standard FFT searches. Additionally, RFI may have contributed to their non-detection. Nevertheless, the detection of $\sim$ 50\% (15 out of 34) of these pulsars in FFT searches suggests that RFI likely did not play a significant role in their non-detection.

The majority of these pulsars were identified near the FFT noise floor (with $\rm S/N_{fft}$ < 6) and in higher harmonic sums. For these pulsars, the mean value of $\rm S/N_{fft}$ is only 5.9. However, after folding, it increases to 9.8, representing a roughly 70\% boost in the S/N.  A similar outcome was observed during the reprocessing of the entire PMPS survey presented in \citet{sengar_23}, where these 15 pulsars were detected blindly. However, the original discovery plots of 10 of them clearly resembled a genuine pulsar detection, therefore they are included as independent discoveries in the PMPS reprocessing \citep{sengar_23} (they were originally discovered for the first time in this work). This underscores that our search techniques also excels in discovering pulsars that are only detectable near the FFT noise floor and were overlooked in previous processings.

\section{Statistical analysis} 
\label{sec:statistical analysis}

\subsection{Period, DM and S/N distribution}
\label{subsec:p_dm_sn}

\begin{figure*}
\centering
\includegraphics[width=0.7\textwidth]{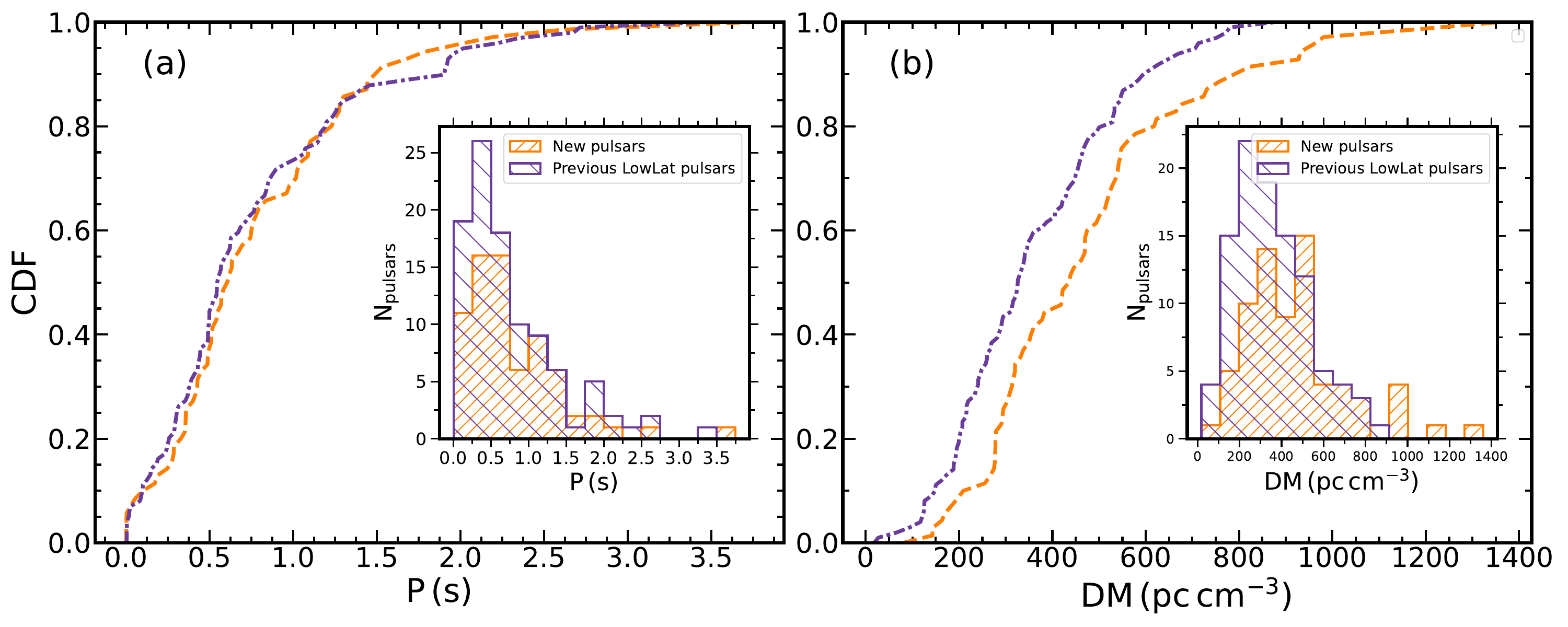}
\includegraphics[width=0.7\textwidth]{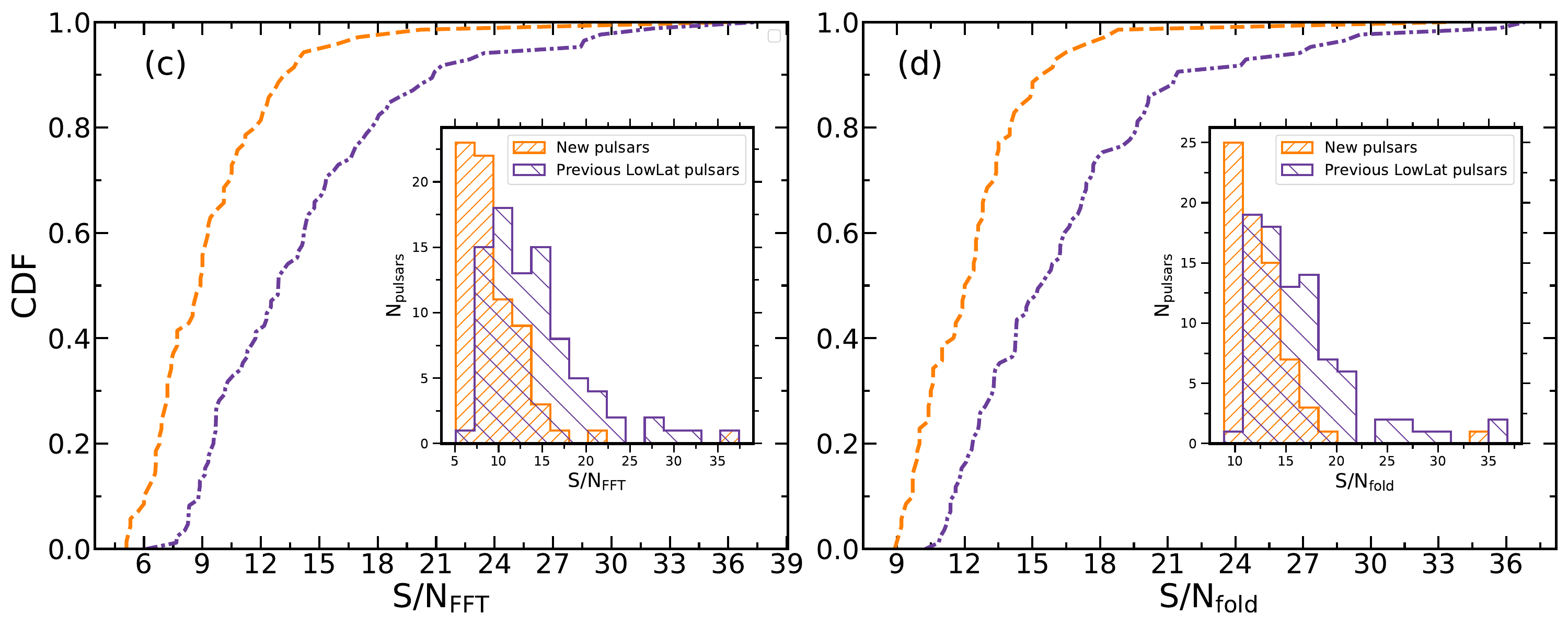}
\caption{ Spin period, DM, $\rm S/N_{FFT}$, $\rm S/N_{fold}$ distribution of previously discovered HTRU-S LowLat pulsar (violet) and new pulsar sample (red) reported in this paper. Each plot shows the CDF of each distribution and shows inside each plot are histograms of the corresponding distribution.}
\label{fig:statistical comparision}
\end{figure*}

A significant increase in new pulsar discoveries by $\sim 75 \%$ as compared to the previous pulsars found in the HTRU-S LowLAT has provided us with a comparable sample of pulsars for evaluating whether our pipeline is identifying pulsars that differ statistically from the previous sample of 100 pulsars reported by \citet{cherry15} and \citet{cameron2020high} in terms of their spin periods, DMs, and signal-to-noise ratios (S/Ns). In Figure~\ref{fig:statistical comparision}(a), we present the cumulative distribution of spin periods for both the previous and newly discovered HTRU-S LowLat pulsars. The distributions overlap, indicating no apparent difference in the distribution of pulsar periods. This observation is supported by a two-sample Kolmogorov-Smirnov test (KS test), a robust non-parametric method for comparing two samples, which lends strong evidence in favour of the null hypothesis, signifying that spin period distribution of the previous HTRU-S LowLat pulsars and the new pulsar sample are statistically indistinguishable.

However, in Figure~\ref{fig:statistical comparision}(b), it is evident that the DM distributions exhibit statistically significant differences, a conclusion corroborated by the KS test with a confidence level of 99.99\% and a $p-$value of $\sim$0.009. The mean and median DM for the 100 pulsars from \citet{cherry15} and \citet{cameron2020high} are 355 and 326 $\rm pc \, cm^{-3}$ respectively. In contrast, for the new pulsar sample, these values are considerably higher at 480 and 436 $\rm pc \, cm^{-3}$, indicating the sensitivity of the reprocessing towards higher DM pulsars.  Notably, the previous sample of pulsars contains only one pulsar, J1731$-$3322, with a DM greater than 800 $\rm pc \, cm^{-3}$ (currently, only 1.9\% of known pulsars have DM values exceeding 800 $\rm pc \, cm^{-3}$). Conversely, our new pulsar sample includes seven pulsars with DM values surpassing this threshold, further affirming that our search pipeline is probing pulsars with higher dispersion measures.

 Figures~\ref{fig:statistical comparision}(c) and (d) illustrate the cumulative distribution of the $\rm S/N_{fft}$ and  $\rm S/N_{fold}$ for the two sets of pulsars discovered in the previous and current analyses. The KS test indicates a significant difference in both of these distributions, with both having confidence levels exceeding 99.99\% with a $p-$value of $\sim$1e-07. Specifically, the median of $\rm S/N_{fft}$ and  $\rm S/N_{fold}$  for the 100 pulsars identified in previous analyses is approximately 12.9 and 15.5, respectively. However, for the new pulsar sample the median $\rm S/N_{fft}$ and  $\rm S/N_{fold}$  are 8.9 and 12 respectively, suggesting that our search technique is targeting pulsars near the detection threshold of the survey. This disparity may be attributed to the previous searches predominantly favoring the detection of brighter pulsars, primarily by selecting candidates above a certain S/N threshold. However, further investigation into the survey reveals underlying factors contributing to this discrepancy which we will discuss in Section \ref{sec:why_so_missed}.

\begin{table}
\centering
\caption{Parameters listed in the table are the distances, $D$  of the pulsars using NE2001 and YMW16 electron density models, the minimum flux densities $(S_{\rm min,1400})$ of pulsars assuming the discovery position is the beam centre, $\rm W_{50}$ is the pulse width at 50\% of the discovery pulse profiles. Derived luminosity, $L_{\rm min, 1400}$ corresponds to $S_{\rm min,1400}$ for both NE2001 and YMW16 models. }
\label{table:pulsar_distance_table}
\begin{tabular}{lllllll}

\Xhline{1\arrayrulewidth}
PSR name &$D$& $S_{\rm min,1400}$ & $\rm W_{50}$ &\multicolumn{2}{l}{ $L_{1400}$ }  \\
 &(kpc)&$({\rm mJy})$& (ms) &\multicolumn{2}{l}{$(\rm mJy \, kpc^{2})$}   \\ 
\Xhline{1\arrayrulewidth}
J1136$-$64&7.4, 3.4&0.07(1)&18.1$\pm$1.8&3.6&0.8\\
J1306$-$60&6.7, 11.2&0.09(1)&156.1$\pm$25.0&4.1&11.4\\
J1310$-$63&11.5, 12.0&0.06(1)&6.1$\pm$1.2&7.8&8.5\\
J1325$-$6253&5.4, 6.4&0.11(8)&1.7$\pm$0.3&3.2&4.5\\
J1333$-$61&9.0, 12.0&0.07(2)&21.0$\pm$1.9&5.8&10.3\\
J1348$-$62&13.2, 12.8&0.14(1)&144.6$\pm$16.0&24.6&23.1\\
J1406$-$59&5.2, 5.5&0.05(2)&23.8$\pm$3.2&1.3&1.4\\ 
J1423$-$62&7.1, 6.5&0.05(1)&66.1$\pm$10.0&2.4&2.1\\
J1437$-$62&6.8, 6.4&0.06(2)&16.4$\pm$0.0&2.6&2.3\\
J1449$-$63&2.2, 1.6&0.18(1)&2.2$\pm$0.4&0.9&0.5\\
J1518$-$60&8.7, 15.1&0.10(1)&23.7$\pm$2.2&7.7&23.4\\
J1521$-$57&4.2, 4.5&0.07(2)&5.8$\pm$0.9&1.2&1.3\\
J1549$-$5337&3.9, 3.5&0.24(1)&0.4$\pm$6.0&3.6&3.0\\
J1548$-$55&6.4, 6.5&0.08(1)&42.5$\pm$0.026&3.5&3.6\\
J1555$-$53&9.6, 6.3&0.17(1)&102.6$\pm$13.0&15.7&6.8\\
J1559$-$55&5.7, 6.4&0.04(1)&18.7$\pm$2.0&1.3&1.6\\ 
J1600$-$49&9.6, 7.1&0.10(1)&23.7$\pm$3.2&9.4&5.2\\
J1603$-$54&7.2, 6.3&0.07(2)&20.2$\pm$3.5&3.6&2.7\\
J1605$-$52&7.2, 4.9&0.07(1)&31.3$\pm$2.9&3.5&1.6\\   
J1613$-$51&10.4, 7.6&0.10(1)&114.1$\pm$16.0&10.9&5.9\\
J1628$-$46&7.3, 11.8&0.07(1)&20.0$\pm$3.0&4.0&10.2\\
J1631$-$47&8.0, 6.8&0.10(1)&51.0$\pm$6.0&6.9&4.9\\
J1632$-$49&9.2, 7.1&0.17(1)&67.2$\pm$8.0&14.2&8.6\\
J1634$-$49&6.6, 7.1&0.10(1)&9.7$\pm$0.8&4.0&4.7\\
J1635$-$46&6.1, 5.0&0.12(1)&31.1$\pm$2.9&4.5&3.0\\
J1638$-$47&17.0, 10.0&0.21(1)&141.7$\pm$15.0&60.8&20.9\\
J1639$-$46&9.3, 5.8&0.09(1)&23.6$\pm$4.0&7.4&2.9\\
J1641$-$49&10.2, 18.4&0.10(1)&47.0$\pm$6.0&10.7&35.2\\
J1647$-$49&9.8, 19.0&0.09(1)&9.6$\pm$1.3&8.5&32.0\\
J1652$-$4237&13.0, 25.0&0.20(2)&92.6$\pm$6.0&34.9&128.6\\
J1651$-$46&6.5, 11.1&0.16(1)&60.0$\pm$13.0&6.6&19.8\\
J1655$-$40&6.7, 18.1&0.10(1)&20.6$\pm$2.7&4.8&34.9\\
J1700$-$39&7.4, 16.7&0.06(1)&53.8$\pm$7.0&3.3&16.7\\
J1708$-$38&6.9, 14.2&0.07(1)&21.3$\pm$3.1&3.2&13.6\\   
J1755$-$22&4.3, 4.3&0.06(1)&17.8$\pm$13.0& 1.1\\
J1710$-$3946&12.4, 7.1&0.14(1)&108.3$\pm$12.0&21.5&7.0\\ 
J1717$-$41&5.6, 12.1&0.10(1)&23.2$\pm$2.3&3.1&14.7\\
J1719$-$36&6.5, 4.6&0.11(1)&19.5$\pm$2.2&4.8&2.4\\  
J1723$-$40&4.3, 4.2&0.05(1)&60.1$\pm$7.1&0.9&0.8\\ 
J1735$-$28&4.4, 8.0&0.09(1)&18.0$\pm$2.2&1.8&5.7\\ 
J1735$-$33&2.8, 3.0&0.07(1)&27.0$\pm$3.2&0.5&0.6\\ 
J1738$-$33&3.9, 3.9&0.08(1)&8.4$\pm$1.6&1.2&1.2\\
J1739$-$26&6.0, 16.1&0.07(1)&13.1$\pm$2.3&2.4&17.0\\
J1756$-$25&3.5, 3.2&0.06(1)&7.5$\pm$1.7&0.7&0.6\\     
J1757$-$26&7.0, 9.3&0.32(2)&125.0$\pm$17.0&15.7&27.8\\
J1758$-$24&10.6, 4.6&0.18(1)&48.7$\pm$5.0&20.0&3.8\\
J1758$-$25&5.5, 4.3&0.09(1)&14.6$\pm$1.9&2.8&1.6\\
J1806$-$2133&12.2, 6.2&0.23(2)&84.4$\pm$10.0&34.1&8.7\\
J1808$-$14&5.2, 8.4&0.06(2)&19.8$\pm$2.9&1.6&4.4\\
J1808$-$19&11.6, 6.3&0.50(3)&24.6$\pm$2.2&66.5&19.7\\
J1809$-$20&6.6, 4.5&0.17(1)&7.4$\pm$1.0&7.6&3.5\\
J1812$-$12&4.9, 5.9&0.050(1)&22.5$\pm$2.6&1.2&1.7\\ 
J1813$-$14&6.2, 12.2&0.08(1)&20.2$\pm$2.3&3.2&12.2\\
J1814$-$1845&6.7, 4.7&0.42(1)&64.5$\pm$3.0&18.8&9.6\\
J1817$-$19&3.9, 3.8&0.14(1)&38.8$\pm$5.0&2.1&2.0\\
J1820$-$19&3.2, 3.2&0.17(1)&0.60$\pm$0.04&1.7&1.8\\ 
J1820$-$20&5.6, 9.9&0.17(1)&0.50$\pm$0.03&5.4&16.8\\ 
J1823$-$11&8.3, 7.0&0.30(2)&82.7$\pm$11.0&19.5&13.7\\
J1830$-$09&4.6, 4.1&0.10(1)&13.4$\pm$1.3&2.3&1.8\\
J1830$-$14&5.1, 5.6&0.07(1)&20.0$\pm$4.0&1.8&2.2\\
\hline
\Xhline{1\arrayrulewidth}
\end{tabular}
\end{table}

\begin{table}
\caption*{\textbf{Continued}}
\label{table:pulsars_table}
\centering
\begin{tabular}{lllllll}
\Xhline{1\arrayrulewidth}
PSR name &$D$& $S_{\rm min,1400}$ & $\rm W_{50}$ &\multicolumn{2}{l}{ $L_{1400}$ }  \\
 &(kpc)&$(\rm mJy)$& (ms) &\multicolumn{2}{l}{$(\rm mJy \, kpc^{2})$}   \\ 
\Xhline{1\arrayrulewidth}
J1833$-$05&6.7, 9.3&0.09(1)&31.6$\pm$5.0&3.9&7.5\\
J1835$-$09&6.7, 5.0&0.20(2)&119.5$\pm$14.0&9.1&5.0\\
J1837$-$10&6.7, 11.6&0.07(1)&35.5$\pm$4.3&2.9&8.8\\ 
J1840$-$09&3.5, 3.7&0.26(2)&1.7$\pm$0.32&3.3&3.7\\
J1841$-$05&5.3, 4.1&0.06(1)&26.8$\pm$2.7&1.8&1.1\\ 
J1842$-$06&7.7, 6.8&0.06(1)&12.8$\pm$2.1&3.8&2.9\\
J1844$-$02&5.9, 4.5&0.07(1)&24.1$\pm$3.5&2.4&1.4\\ 
J1844$-$09&8.2, 18.5&0.05(1)&15.1$\pm$2.7&3.7&18.8\\
J1846$-$05&7.7, 9.0&0.09(1)&32.9$\pm$4.0&5.2&7.1\\
J1847$-$05&3.3, 3.7&0.04(1)&39.2$\pm$2.0&0.4&0.5\\ 
J1854$-$05&6.0, 8.4&0.04(1)&22.6$\pm$5.0&1.4&2.8\\

\Xhline{1\arrayrulewidth}
\end{tabular}
\end{table}

\subsection{Flux density distribution}
\label{Flux_density_calculations}

In an ideal scenario, pulsar flux densities are determined through a two-step process: first, by obtaining an optimal position of the pulsar, and then by calibrating the data using the radiometer equation or through comparision to calibration signals that have been referred to other sources of known flux density. However, in cases where calibrated data and the precise pulsar positions are unavailable, we opted to estimate the minimum flux densities at 1400 MHz ($S_{\rm min, 1400}$) by assuming the discovery position as the center of the observing beam. For this estimation, we utilized the pulsar's discovery profile shown in Figure \ref{fig:pulse profiles}. Our method for flux density calibration is similar to the one described in \cite{gitika_23} which employs the radiometer equation and assumes that the pulsar's off-pulse baseline rms is accurately represented by the sum of the system's equivalent flux density and the Galactic sky temperature, all divided by the antenna gain. The MB receiver has different gain values for its distinct rings: 0.581 K\,Jy$^{-1}$ for the outer rings, 0.690 K\,Jy$^{-1}$ for the inner rings, and 0.735 K\,Jy$^{-1}$ for the center ring. Therefore, for the flux density estimation, we selected the appropriate gain value corresponding to beam in which the pulsar was identified. The background sky temperature at the location of the pulsar was obtained from all-sky catalogue at 408 MHz \citep{haslam82} and scaled at 1400 GHz as $\nu^{-2.6}$. We have provided the $S_{\rm min, 1400}$ values in Table \ref{table:pulsar_distance_table}.

\begin{figure}
\vspace{-0.5cm}
\begin{center}
\includegraphics[width=0.80\columnwidth,angle=0]{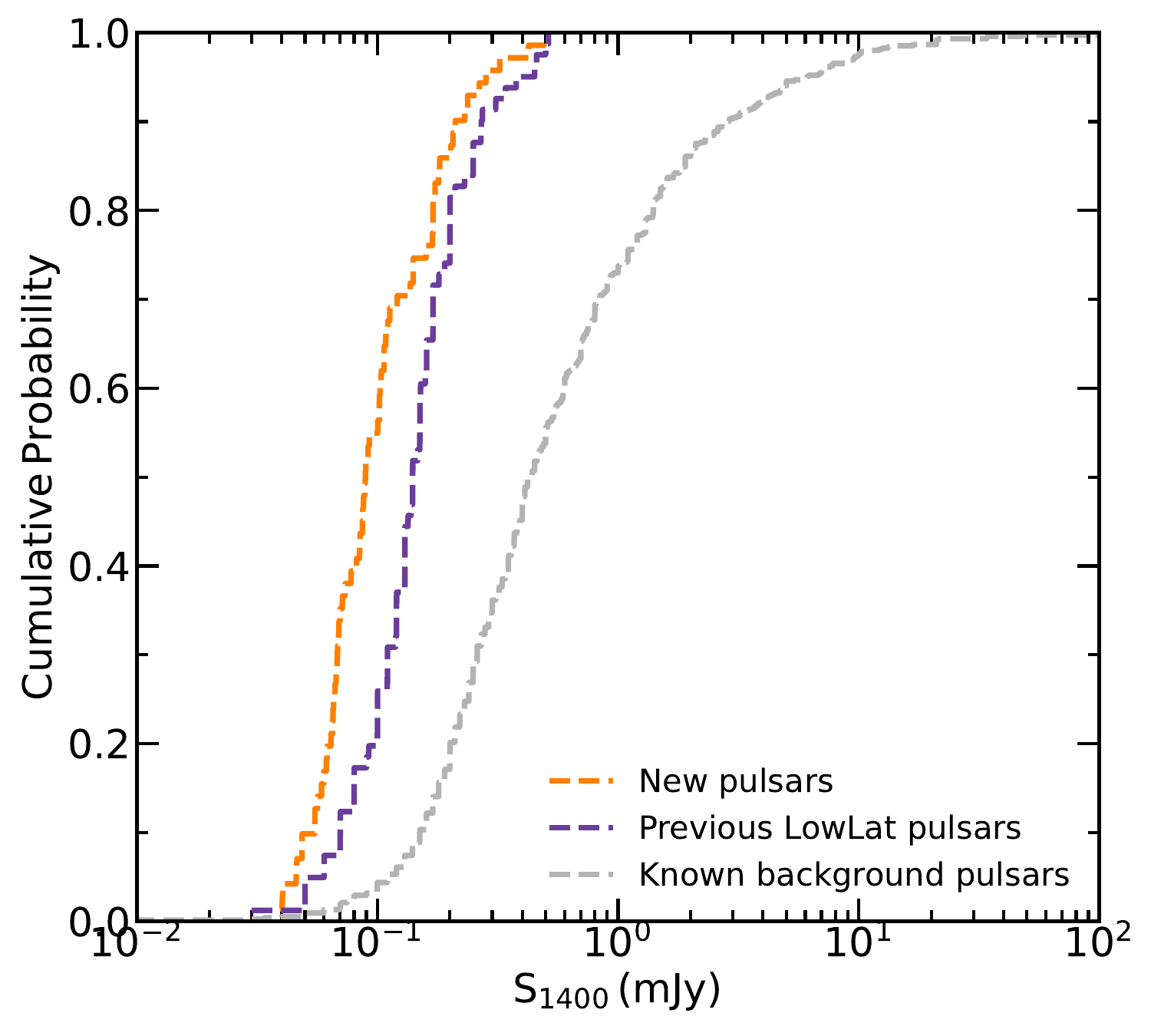}
\caption{The cumulative distribution function of flux densities of pulsars in the HTRU-S LowLat region. The background pulsars are shown in grey, the previous HTRU-S LowLat puslars are shown in violet and the new pulsars reported in this paper are in orange.}
\label{fig:s1400}
\end{center}
\end{figure}

As detailed in Section \ref{subsec:p_dm_sn}, the S/N distribution of the new sample differs significantly from the previous 100 HTRU-S LowLat pulsars. Therefore, it is interesting to investigate whether the flux density of the new sample is drawn from a distinct population compared to the previous HTRU-S LowLat pulsars. For comparing $S_{\rm min, 1400}$ of the new pulsar sample against the previous HTRU-S LowLat pulsars, in Figure \ref{fig:s1400} we show the cumulative probabilities of the new sample, previous HTRU-S LowLat pulsars and the background pulsar population. Although, the previous HTRU-S LowLat pulsars are noticeably fainter than the background pulsar population \citep{cameron2020high}, a two-sample KS test analysis reveals that the new sample probes even fainter sources with a significance greater than 99.99\% ($p-\mathrm{value} < 10^{-5}$).

However, it's important to note that the true position of the new pulsar sample could potentially alter the flux densities of the new pulsar sample, as the flux density would increase according to Equation \ref{eq:s1400_factor}. To place practical limits on the flux densities of the new pulsar sample, we determined the average positional offset of the previous 88 HTRU-S LowLat pulsars from their discovery observations, relative to the timing positions reported in \cite{cherry15} and \cite{cameron2020high}. This offset was found to be 3.94 arc minutes from the beam center. When comparing the newly obtained flux densities of the new pulsar sample with this offset, the KS test analysis yields a $p$-value of 0.014, providing evidence that the new sample is probing relatively fainter pulsars.

\subsection{Distance and Luminosity}
\label{subsec:distances_luminosities}

\begin{figure}
\centering
\includegraphics[width=0.45\textwidth]{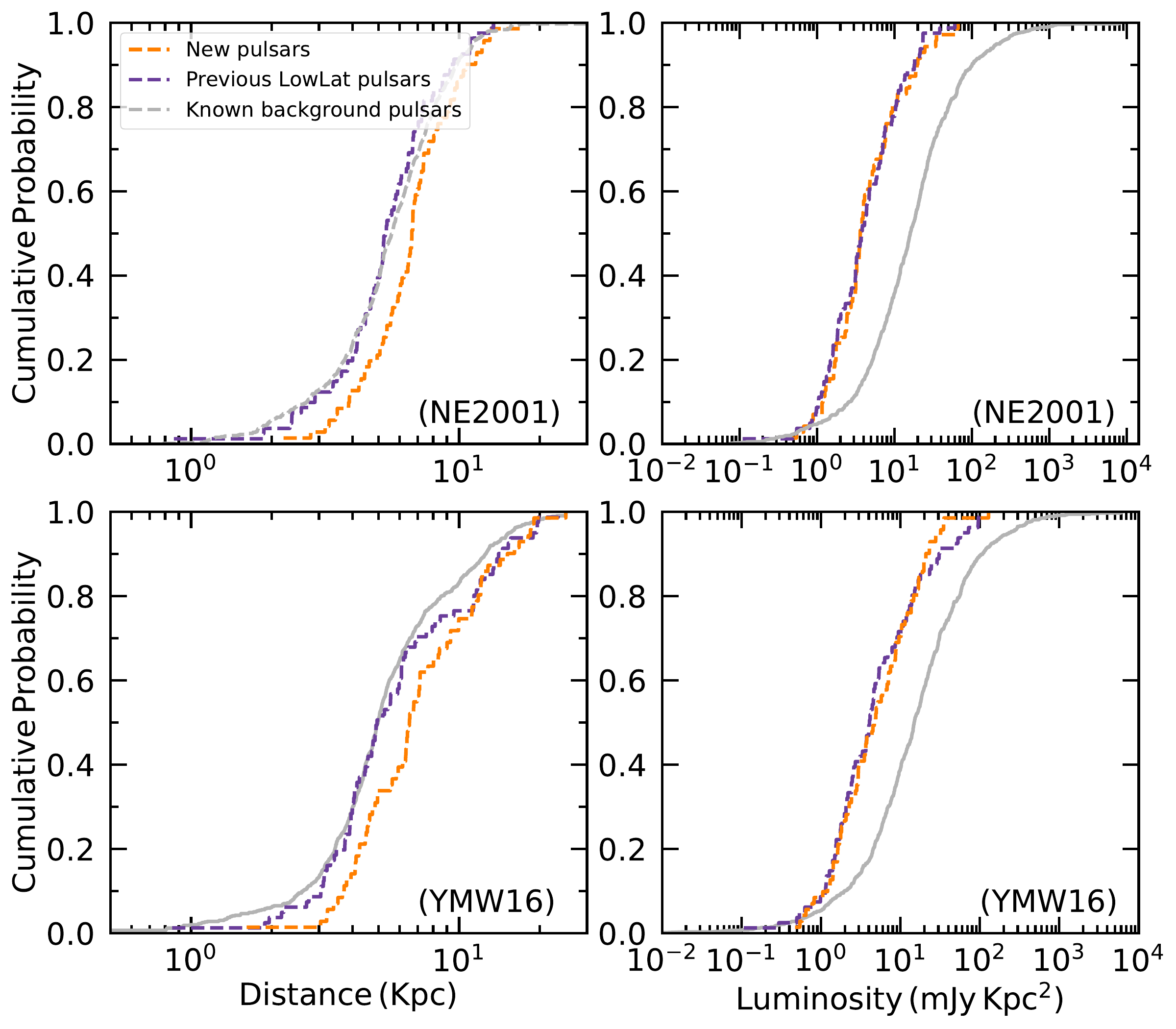}
\caption{The cumulative distribution functions (CDFs) of distance and luminosity of the pulsars at 1400 MHz. The known background pulsars are shown grey, previous HTRU-S LowLat pulsars in violet and 71 new HTRU-S LowLat pulsars in orange. On the left are the CDFs of distance distribution of each population  under both the NE2001 (top) and YMW16 (bottom) electron density models. The corresponding CDFs for luminosity distribution are shown on right.}

\label{fig:distance_lum}
\end{figure}

As discussed in \ref{subsec:p_dm_sn}, the new sample of HTRU-S LowLat pulsars is probing statistically higher DM pulsars. This prompted us to investigate whether the DM-derived distances of these pulsars are similar or different. We estimate the distances of all pulsars using the NE2001 \citep{ne2001} and YMW16 \citep{ymw16} electron density models, and for each model, luminosities were derived using the inverse square relation, $L = S_{1400} \times D^{2}$ where $D$ is model dependent pulsar distance. The corresponding value of $S_{1400}$ and $ L_{1400}$ for each pulsar are listed in Table \ref{table:pulsar_distance_table}.

In Figure \ref{fig:distance_lum}, we show the distance and luminosity comparison of the new 71-pulsar sample, previous HTRU-S LowLat pulsars and the known background population. When comparing the distance distribution of previous HTRU-S LowLat pulsars against the known background pulsars (shown in the left panel of Figure \ref{fig:distance_lum}), the distance distributions of both populations are similar and consistent with the results obtained by \citet{cameron2020high}. However, as confirmed by the KS test with $p$-value of 0.007 and 0.009, the new pulsar sample is probing more distant pulsars with a median distance of 6.7 kpc and 6.5 kpc, for both the NE2001 and YMW16 electron density models, respectively. This distance is 20-35\% greater than the previous HTRU-S LowLat and background pulsar population. However, the DM inferred distances are not well constrained \citep[e.g.,][]{deller_19}, therefore they should be interpreted with caution.

When comparing luminosities, the KS test indicates that both the previous and new HTRU-S LowLat pulsars are less luminous than the known background pulsars ($p$-value $<10^{-5}$). However, they exhibit similar luminosities when compared to each other. This similarity in luminosity between previous and new HTRU-S LowLat pulsars is likely due to the new pulsars having lower flux densities but larger distances. However, it is important to note that due to uncertainties in both the distance and flux density measurements, these luminosity estimates should also be interpreted with caution.


\begin{figure*}
\centering
\includegraphics[width=0.80\textwidth]{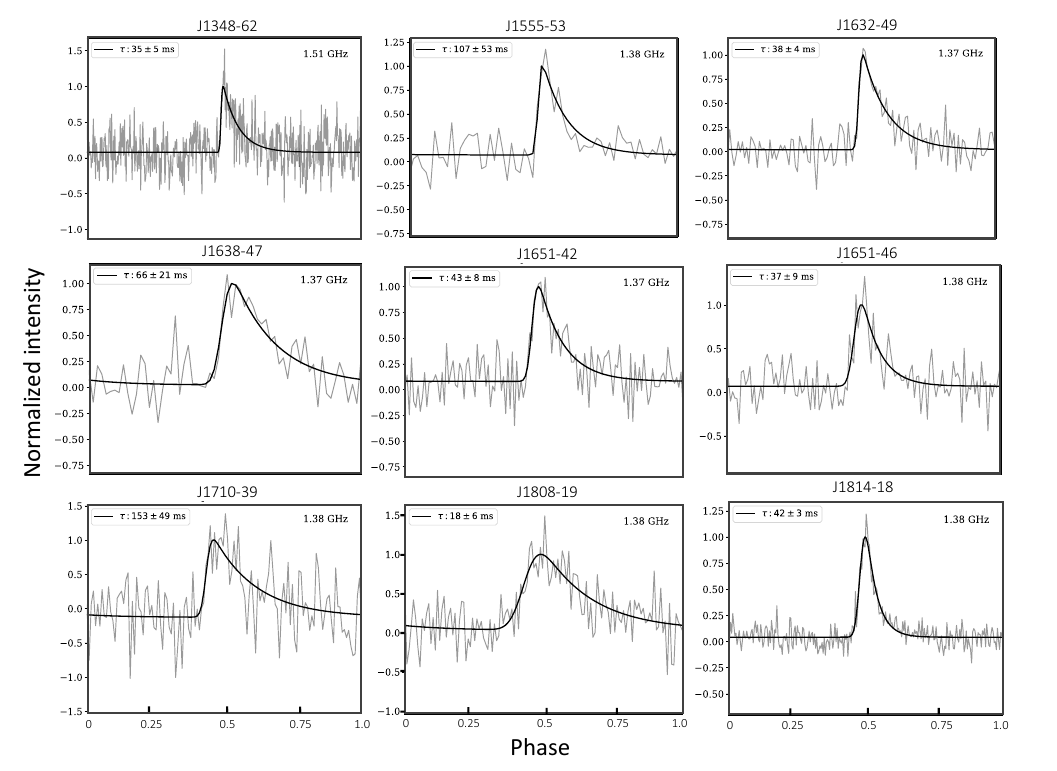}
\caption{The scatter-broadened pulse profiles (in grey) of newly discovered HTRU-S LowLat pulsars. The solid black line represents the best-fit pulse broadening function discussed. In each plot, the value of scattering timescale, $\tau_{\rm sc}$ in $\rm ms$ and the effective central frequencies are listed.}

\label{fig:scattering_plot}
\end{figure*}

\subsection{Scattering }
\label{subsec:scattering_analysis}

The presence of intervening plasma in the Interstellar Medium (ISM) leads to the scattering of a pulsar's pulse profiles. This scattering effect is dependent upon the frequency, being more pronounced at lower frequencies and scaling inversely with the fourth power of the frequency, i.e. $\nu^{-4}$. Analyzing these scattering properties provides crucial insights into the ISM's structure. It aids in refining models of electron density and mitigating perturbations in Times-of-Arrival (TOAs) by disentangling scattering effects from the intrinsic pulse profile of the pulsar \citep{lenati_17}.

Among our recently discovered 71 LowLat pulsars, nine exhibit visible scattering tail. This characteristic enabled us to conduct a scattering analysis using a software package \texttt{SCAMP-I}\footnote{\url{https://github.com/pulsarise/SCAMP-I}} which models the pulsar's intrinsic Gaussian pulse profile convolved with an exponential function, $\tau_{\rm scat}^{-1}e^{-1/\tau_{\rm scat}}$, where $\tau_{\rm scat}$ denotes the scattering timescale. This timescale scales with frequency as $\tau_{\rm scat} \propto \nu^{-\alpha_{\rm scat}}$.

Due to low S/Ns of these pulsars, employing sub-banded profiles for fitting the model and determining $\tau_{\rm scat}$ at different frequencies and scattering indices was not feasible. Therefore, we integrated the profiles across the entire frequency band and obtained their $\tau_{\rm scat}$. The scattering model fit to the pulse profiles from the original discoveries is shown in Figure \ref{fig:scattering_plot}. The obtained scattering timescale values, $\tau_{\rm scat}$, have been normalized to 1 GHz values and are presented in Table \ref{table:tau}. Additionally, we show the computed $\tau_{\rm scat}$ using the NE2001 electron density model and scaling relation from \cite{krishna15} at 1.0 GHz, assuming a scattering index, $\alpha_{\rm scat}$, of 4.0.

\begin{figure}
\centering
\includegraphics[width=0.4\textwidth]{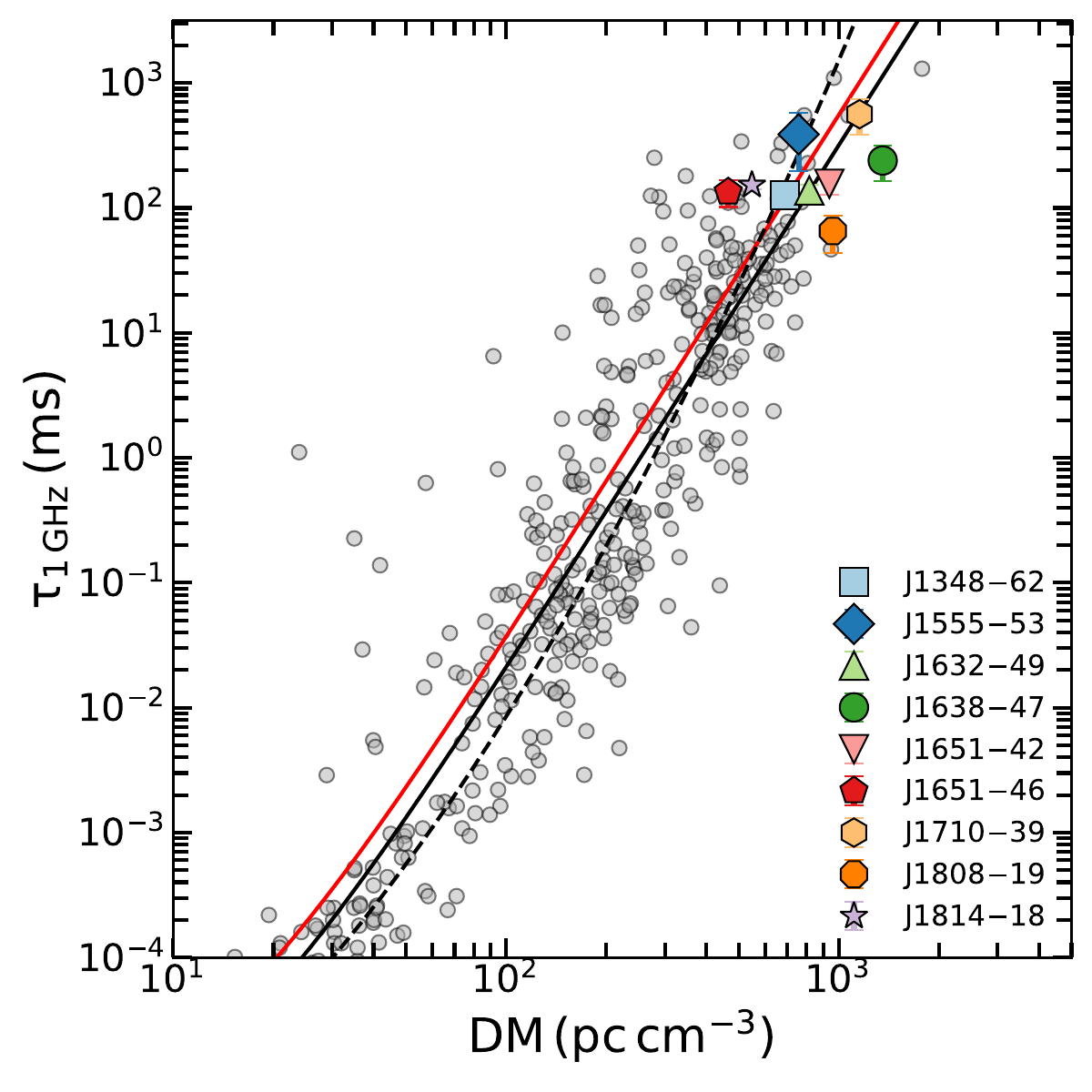}
\caption{DM vs. scattering timescales of the pulsars presented in this work and reported in \texttt{PSRCAT} with the measurements of $\tau_{\mathrm{sc}}$ (grey). The dotted black line is the relation from \citet{bhat_04}, and the solid red and solid black lines represent the relation between $\tau$ and DM obtained by \citet{krishna15}, scaled at 1 GHz and assuming a scattering spectral index ($\alpha_{\mathrm{sc}}$) of $-3.5$ and $-4.0$, respectively.}
\label{fig:tau_dm}
\end{figure}

In figure \ref{fig:tau_dm}, we plot the DM versus $\tau_{\rm sc}$ of these pulsars along with the $\tau_{\rm sc}$ of the known pulsars in the \texttt{PSRCAT} and clearly the $\tau_{\rm scat}$ of the new pulsars probe the extreme end of $\tau_{\rm scat}$ distribution. Notably, the values of $\tau_{\rm scat}$ derived from NE2001 and empirical relations from \cite{krishna15} and \cite{bhat_04} substantially deviate from our measured values for most pulsars. A plausible explanation for these disparities could be the presence of unaccounted foreground structures, such as H II regions in the NE2001 electron density models.

\begin{table}
\caption[Pulse scattering parameters.]{Pulse scattering parameters obtained by using the model briefly explained in Section \ref{subsec:scattering_analysis}. The values of scattering timescale, $\tau_{\rm scat}$ in each case is obtained at the effective central frequency of 1.38 GHz and have been normalized to 1 GHz. Scattering timescales at 1 GHz predicted by the NE2001 model ($\tau_{\rm scat, \, NE2001}$) and the scaling relation obtained by \citet{krishna15} ($\tau_{\rm  kmn}$) assuming a scattering index, $\alpha_{\rm scat}=4.0$ are given in the last two columns.}
~\\

\centering
\begin{tabular}{ccccc}
\hline
PSR Name &  $\rm DM$ &  $\tau_{\rm scat}$ &  $\tau_{\rm scat, \, NE2001}$ &  $\tau_{\rm kmn}$ \\
 & ($\rm pc \, cm^{-3}$)   &  (ms) &(ms)  &(ms) \\ \hline \hline
J1348$-$62  & 790(30) &126(18)     &  1   &  18.2\\
J1555$-$53  & 785(25) &388(192)   &  54   &  27.3\\
J1632$-$49  & 814(13) &137(14)     &  54   &  37.2\\
J1638$-$47  & 1374(19)&239(76)    &  236   &  313.1\\
J1652$−$4237& 943(14) &156(29)    &  52   &  66.4\\
J1651$-$46  & 487(11) &134(32)     &   5   &  3.5\\
J1710$−$3946& 1198(29)& 562(177)  &  140   &  160.0\\
J1808$-$19  & 969(3)  &65(21)     &   97  &  73.6\\
J1814$−$1845& 534(10) &152(11)     &   27  &  7.0\\ \hline
\label{table:tau}
\end{tabular}
\end{table}

\section{Survey and pipeline efficiency}
\label{sec:pipeline_efficiency}

\subsection{Redetections of known pulsars}
\label{subsec:known_pulsar_redetections}

Through an analysis of 14103 beams in the survey, we identified 792 unique pulsars across 2126 individual beams. These redetections included 86 pulsars among 100 pulsars that were reported as new discoveries by \cite{cherry15} and \cite{cameron2020high} in their first processing of the survey. The remaining 14 previous HTRU pulsars are present in the beams which have not been reprocessed yet. In addition, we also redetected 87 pulsars (including 9 MSPs) in 117 beams which are present in the survey region, but were undetected in the previous processing passes of the survey. To assess the effectiveness of the reprocessing, we conducted a comparison between the detected Signal-to-Noise ratio ($\rm S/N_{\rm detect}=S/N_{\rm fold}$) and the expected Signal-to-Noise ratio ($\rm S/N_{\rm exp}$) for pulsars with an offset, $\theta$, less than half of the Full Width at Half Maximum (FWHM), the value of which is $0.12^{\circ}$ \footnote{In this analysis, the offset, defined as $\theta$=FWHM/2=0.12, is chosen because the response pattern of the MB receiver deviates from a Gaussian approximation beyond the FWHM of the beam. Observations of known pulsars outside this range can lead to inaccurate signal-to-noise ratios (S/N)}. The expected $\rm S/N_{\rm exp}$ of a pulsar can be computed using the radiometer equation:

\begin{align}
\label{eq:radiometer_eq}
S/N_{\rm exp} = S_{\exp}   \dfrac{G\sqrt{n_{\rm{p}} t_{\rm{int}} \Delta f}}{\beta T_{\rm sys}} \sqrt{\dfrac{1-\delta}{\delta}},
\end{align}

where $S_{\rm exp}$ represents the expected equivalent continuum flux density (in Jy) of the pulsar factoring in its degradation due to an offset, $\theta$, from its true position. The duty cycle, $\delta=W_{50}/P$ is the fraction of spin period covered by the observed pulse of the pulsar, $n_{\rm p}$ is the number of polarization, $t_{\rm int}$ in the integration time and $\Delta f$ is the effective bandwidth.  $S_{\rm exp}$ is calculated using:

\begin{equation}
\label{eq:s1400_factor}
      S_{\rm exp} = S_{1400}  \exp \bigg(\dfrac{ -\theta^{2}}{2\sigma^{2}} \bigg) ,
\end{equation}

where $\sigma = \rm FWHM/2.355$ is the width of the beam at $e^{-1/2}$ of beam centre gain. For the MB receiver beam the FWHM is $0.24^{\circ}$, so $\sigma = 0.1^{\circ}$. \par

The other terms in the radiometer equation are related to the telescope properties and observational setup. The value of telescope gain, $G$ ($\rm K \,{Jy^{-1}}$) depends on telescope size and beam configuration of the MB receiver where for the central, six inner and six outer beams the gain values are 0.735 ($\rm K \,{Jy^{-1}}$) , 0.690 ($\rm K \,{Jy^{-1}}$),  and 0.581 ($\rm K \,{Jy^{-1}}$),  respectively \citep{keith10}. The system temperature, $T_{\rm sys}$ is the sum of receiver temperature ($T_{\rm receiver} = 23$\,K) and the mean sky temperature ($T_{\rm sky} = 7.6$\,K) for the sky region covered by HTRU-S LowLat survey. The observations are recorded with an integration length, $t_{\rm int}$= 4320 s, an effective bandwidth, $\Delta f = 340$\,MHz and the number of polarizations summed, $n_{\rm p} = 2$. Finally, $\beta(\geq 1) = 1.16$ is the degradation factor which accounts for the imperfections arise due to digitization of the signal and other effects \citep{handbook04}.\par

\begin{figure}
\centering
\includegraphics[width=0.4\textwidth]{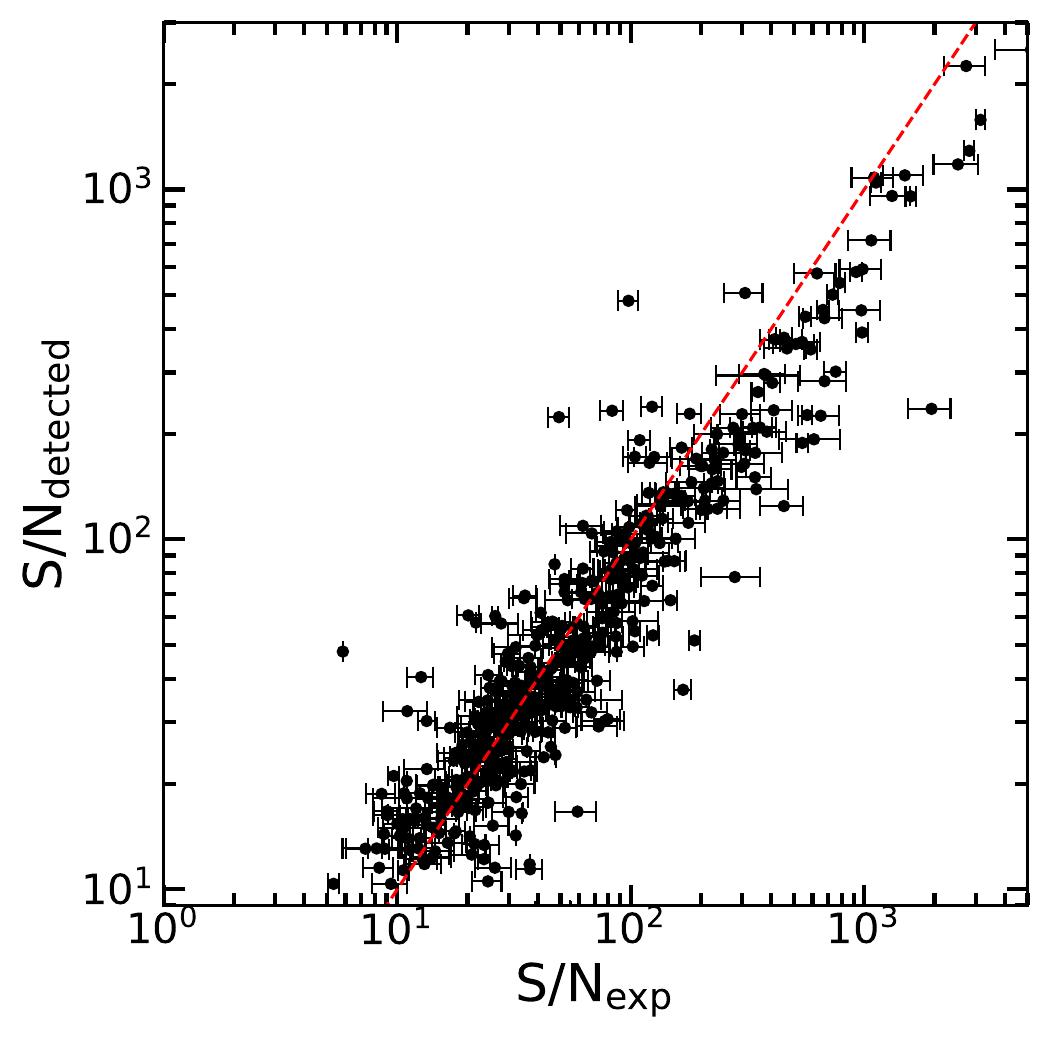}
\caption{The distribution of detected S/N ($\rm S/N_{detected}$) and expected S/N ($\rm S/N_{expected}$) of 467 unique known pulsars. The red dashed line is line of quality (1:1 correlation line). The error bars shown are the uncertainties from \texttt{PSRCAT}. }

\label{fig:sn_exp}
\end{figure}

The distribution of $\rm S/N_{exp}$ in relation to $\rm S/N_{\rm detect}$ is illustrated in Figure \ref{fig:sn_exp}. To ensure the robustness of our analysis, we excluded redetections of pulsars with positional offsets exceeding half of the FWHM of the telescope beam response pattern. This resulted in a selection of 457 unique pulsars for further examination. The majority of detections cluster around the $1$:$1$ line, indicating that the pulsars identified in this reprocessing exhibit the anticipated sensitivity levels. However, the distribution of pulsars around this relationship is not symmetric. Specifically, 279 pulsars (61$\%$) exhibit $\rm S/N_{exp} > S/N_{obs}$. Notably, this trend is more pronounced for pulsars with high $\rm S/N_{obs}$, exceeding 50. Out of the 262 pulsars with $\rm S/N_{obs} > 50$, 194 pulsars (approximately 74$\%$) have $\rm S/N_{\exp}>S/N_{obs} $. Conversely, for pulsars with $\rm S/N_{obs} < 50$, the 1:1 relationship is nearly perfect, with only 51$\%$ of pulsars exceeding $\rm S/N_{\exp}> S/N_{obs}$. These discrepancies may arise from various factors. For instance, the radiometer equation assumes the pulse profile as a top-hat which is not correct and can result in different values of $\rm S/N_{exp}$, in the case of high $\rm S/N_{obs}$ pulsars, the sensitivity of the survey may be compromised due to the 2-bit digitization of the HTRU-S LowLat data. Additionally, it is possible that in some cases, the tendency to publish discovery observations, which typically have the highest $\rm S/N_{obs}$ due to scintillation, leads to higher flux densities in the \texttt{PSRCAT}. The influence of scintillation can also exhibit a relatively large scatter for low DM pulsars.

\subsection{Evaluation of survey yield}
\label{subsec:yield_eval}

Given that we have now reprocessed an equivalent amount of the HTRU-S LowLat data ($\sim 87.3 \%$) as compared to 94$\%$ of data previously processed by \cite{cherry15} ($50\%$) and \cite{cameron2020high}($44\%$), we are in a position to re-evaluate the survey yield and a comprehensive performance of the survey by comparing with the predictions made using population synthesis based studies of the survey. In Table \ref{tab:simulations}, we have listed the pulsar predictions (adjusted for 94 \% of the survey) made from different population studies. The first population-based study of the survey was conducted by \cite{keith10} using \texttt{PSRPOP}\footnote{\url{https://psrpop.sourceforge.net/}}\citep{psrpop_11} and predicted that the HTRU-S LowLat survey should be able to detect a total of 900 normal pulsars ($P>30 \, \rm ms$). Later, \cite{cherry15} conducted a similar study using \texttt{PsrPOPPy}\footnote{\url{https://github.com/samb8s/PsrPOPPy}} software \citep{bates_14} and found that there should be a total of 960 normal pulsar detections in the survey.

\begin{table}

\caption{Expected and actual pulsar detections in the HTRU-S LowLat survey. The table provides an overview of the anticipated and observed pulsar detections from four distinct simulations for the HTRU-S LowLat survey. The values have been adjusted to account for 94\% of the survey data. Pulsars are categorized into two groups: ``normal pulsars" ($\rm P>30 \, ms$), and ``MSPs" with ($\rm P < 30 \, ms$). The numbers in rows without parentheses or outside parentheses represent expected number of pulsars detectable in 94\% of the HTRU-S LowLat survey. Inside parentheses, the first number represents the total number of pulsars detected in the processing, including new discoveries, while the number after the slash ("/") indicates only new pulsar discoveries.}
\centering

\begin{tabular}{lccc}
\hline
Author & Normal & MSPs  \\
\hline
\citet{keith10}  & 900  & 48  \\
\citet{levin013} & -- & 64 \\
\citet{cherry15} \text{ \& }    & 960 (723/92) & 40 (26/8) \\ 
\citet{cameron2020high}   &   &  \\ 
This Work$^{*}$ & 890$\pm$ 27 (879/87) & 40$\pm$4 (43/7) \\
\hline
\end{tabular}
\label{tab:simulations}

\end{table}

To further refine these predictions, we also conducted simulations using \texttt{PsrPOPPy} and found that HTRU-S LowLat survey should detect approximately 890 normal pulsars and 40 MSPs in 94 \% of survey area. However, it is important to note that these numbers carry inherent uncertainties due to the variability in detection methods and the assumptions used in simulations. \cite{cameron2020high} reported a total of 723 normal pulsar detections from 94$\%$ processing of the survey (including 50$\%$ of the survey processed by \citet{cherry15} ), which are 20-25\% fewer detections than  predictions for the 94 \% of the survey. To maintain consistency with the previous processing of the survey, in 87.3$\%$ of the survey $\sim$ 890 normal pulsars should be detected. As outlined in Section \ref{subsec:known_pulsar_redetections}, aside from the 86 redetections of previous HTRU-S LowLat pulsars among which 78 are normal pulsars, there have been 706 redetections of known background pulsars, with 670 falling under the category of known normal pulsars. Apart from this, 23 new pulsars were also found in two parallel reprocessings of the survey (see Section \ref{sec:htru_survey_intro}) using different search algorithms, including one MSP. Therefore, the total new discoveries in HTRU-S LowLat survey are 94, among which 87 are normal pulsars and 7 MSPs. This brings the overall normal pulsars detection count to 835. 

In the remaining 7\% of the data there are 14 previously detected LowLat pulsars and 23 known normal pulsars, all of which were redetected by both \cite{cherry15} and \cite{cameron2020high}. If this 7\% segment of the survey undergoes reprocessing, we can anticipate redetection of 37 normal pulsars. Moreover, extrapolating from this, we project the discovery of an additional 7 new normal pulsars. This combined total of 44 pulsars would bring the tally of normal pulsars from 94\% of the survey to 879, aligning closely with predictions from simulations, and only slightly below the predictions made by \cite{cherry15} by a factor of $\sim 8\%$. Conversely, it falls short of the predictions by \cite{keith10} by only $\sim 2.5 \%$ and 1.1 \% when compared with the population results in this work. \par

\begin{table*}

\caption{Known millisecond pulsars that were missed in previous analyses of the HTRU-S LowLat survey, but redetected in this reprocessing. The exact reasons for their oversight in previous analyses remain unclear, however it is likely due to human error during candidate inspection or suboptimal data downsampling. The table lists the pointing/beam in which the pulsar was redetected, the offset (angular distance) from the beam centre to the coordinates in \texttt{psrcat}, $\rm P_{cat}$ and $\rm DM_{cat}$ are the period and dispersion measure values from \texttt{PSRCAT}, $\rm S_{exp}$ is the theoretical SN of the pulsar calculated from the radiometer equation (assuming a $5\%$ duty cycle), and $\rm nh$ are the number of harmonic sums which resulted in a spectral signal-to-noise-ratio ($\rm SN_{FFT}$). $\rm SN_{fold}$ is the S/N after folding.} 
\label{table:missing_msps}
\centering
\begin{tabular}{lllllllllllll}
\hline

PSR name & pointing/beam &offset & $\rm P_{cat}$ & $\rm DM_{cat}$ & nh & $\rm SN_{FFT}$ & $\rm S/N_{fold}$ & comments \\

& & $\rm (^{\rm \circ})$ &(ms) &$\rm (pc \,cm^{-3})$ & &  &\\
\hline

J1125$-$6014&2011-12-09-18:06:17/10&0.0978&2.63038&52.951&3&30.4&35.1&Binary pulsar\\
&&&&&&&& \cite{lorimer06}\\

J1337$-$6423&2012-04-13-10:37:54/02&0.096&9.42341&259.9&3&20.9&21.5&Intermediate binary pulsar\\
&&&&&&&& \cite{keith011}\\

J1431$-$6328&2011-04-17-09:02:20/13&2.080&2.7723&228.2&0&22.9&20.5&Highly polarised source\\
&&&&&&&& \cite{kaplan019}\\

J1543$-$5440$^{\dagger}$&2011-10-09-08:55:32/12&0.11&4.313&102.1&2&11.0&13.0 & Binary pulsar\\
&&&&&&&& \cite{padmanabh_23}\\
J1546$-$5925&2011-04-24-10:48:37/13&0.28&7.8&168.3&3&9.2&11.4&\cite{mickaliger012}\\
J1552$-$4937&2012-01-19-17:59:08/05&0.10&6.28431&114.6&2&19.7&18.8&\cite{faulkner04}\\
J1652$-$4838&2012-12-13-20:55:26/03&0.45&3.78512&187.8&1&59.7&52.3&Binary MSP\\
&&&&&&&& \cite{knispel13}\\
J1725$-$3853&2011-10-10-06:26:42/07&0.13&4.79182&158.2&1&12.6&13.4&Isolated MSP\\
&&&&&&&& \cite{mickaliger012}\\

J1748$-$2446C&2013-04-01-15:35:14/03&0.44&8.4361&237.0&0&15.9&15.8&\cite{lyne2000}\\

\hline
\end{tabular}

\raggedright{
\small{${^{\dagger}}$ We detected PSR J1543--5440 in 2020, but presumed it as a known pulsar found in HiLat survey. However, upon further review, we recently discovered that it was only listed as a candidate in the HiLat survey and hadn't been subjected to confirmation observation. The pulsar has now been confirmed and published by the MPIfR-MeerKAT Galactic plane survey as PSR J1543$-$5439.\\
}}
\end{table*}  

Apart from the normal pulsar population, three different predictions were also made for the MSP population.
\cite{keith10} and \cite{cherry15} estimated that HTRU-S LowLat should yield 48 and 40 MSP detections, respectively. In our simulations, we also found that 40 MSPs should be detectable. However, \cite{levin013} anticipated a higher MSP yield of 64. Considering the range of estimates (40-64), only 26 MSPs were identified (including 8 new MSPs) in previous processing of the survey, which is 1.5--2.4 times lower than the initial projections. In this work, we have detected a total of 35 previously known MSPs, including 9 MSPs that are present in the survey beams, but eluded their detection in previous processing (see Table \ref{table:missing_msps}). Additionally, 7 more MSPs have also been discovered, resulting in a total count of 42 MSPs in 87.3$\%$ of the survey. We also anticipate one additional MSP from the remaining 7\% of the data, bringing the total to 43 MSPs. This aligns with the lower limit of predictions, but falls about 20 MSPs short of the upper limit which may be due to scattering models that are overly optimistic regarding their impact on the MSP population at 1400 MHz.

\begin{figure}
\centering
\includegraphics[width=0.5\textwidth]{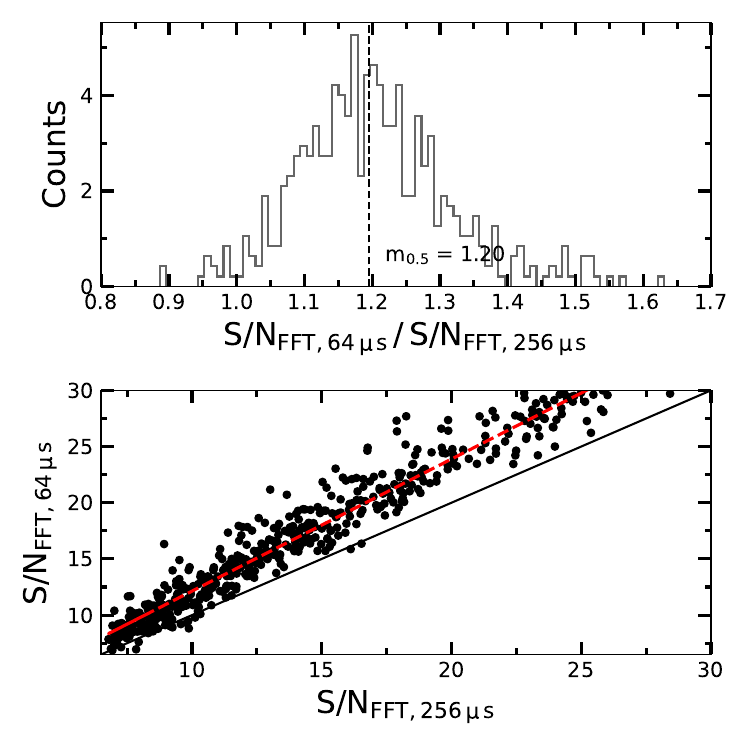}
\caption{Effect of the incorrect down-sampling algorithm on the FFT S/N. Lower panel shows the distribution of detected FFT S/N for known pulsars by down-sampling the data by a factor 4 with the incorrect scaling
factor ($\rm t_{samp} = 256 \mu s$) and with full resolution ($\rm t_{samp} = 64 \mu s$). The dashed red line is the best fit line to the distribution and the solid black line represents 1:1 ratio. It is clearly visible that for full resolution the FFT S/N is consistently higher than the down-sampled data. Upper panel shows the histogram of the same distribution where the median value shows that the full resolution results in effective increase in FFT S/N by $\sim 20 \%$.  }
\label{fig:downsampled_sn}
\end{figure}

\section{Why pulsars were missed in previous processings}
\label{sec:why_so_missed}

 The $\sim70 \%$ increment in survey yield from this latest reprocessing effort raises an intriguing question: why were these pulsars initially not detected? A comprehensive investigation reveals two major reasons, shedding light on the intricacies of the data processing methods employed previously.

\begin{figure}
\centering
\includegraphics[width=0.4\textwidth]{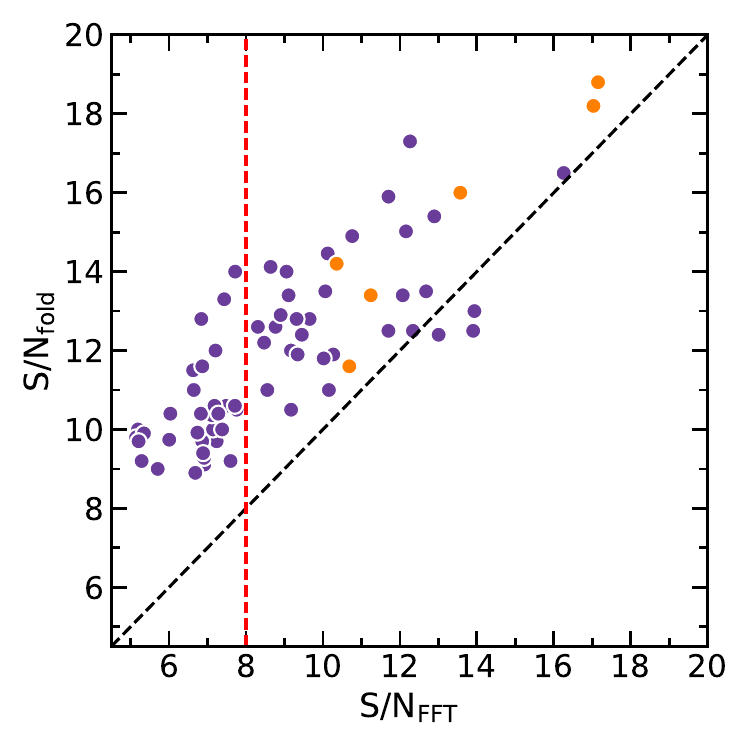}
\caption{$\rm S/N_{fft}$ vs. $\rm S/N_{fold}$ of the 71 newly discovered pulsars in the reprocessing of the HTRU-S LowLat survey. The points in violet correspond to the normal pulsars ($P$>30 ms), while the points in orange are MSPs ($P$<30 ms). The dashed red line divides $\rm S/N_{fft}$ pulsars below and above the false alarm threshold of 8, which was used in previous processing passes of the survey. The black dotted line is 1:1, which indicates that the $\rm S/N_{fold}$ of almost all pulsars is higher than the $\rm S/N_{fft}$. The high S/N (33) detection of PSR J1814$-$18 has been removed to provide clarity for the other data points. }
\label{fig:pulsars_fft_fold_sn}
\end{figure}

\begin{itemize}

    \item In the initial phases of the processing conducted by \citet{cherry15} and \citet{cameron2020high}, the data underwent a four-fold downsampling in time prior to FFT and acceleration searches for full-length 72-m observations with $\rm |a|<1 \, ms^{-2}$. For more extreme acceleration searches, the data were further decimated by a factor of 2 in each step. Upon conducting FFT searches for the complete observations of the known pulsars within the survey after down-sampling, we observed an apparent reduction in FFT signal-to-noise ratio ($\rm S/N_{FFT}$) of approximately 20\% for data down-sampled by a factor of 4. This loss increased to about 25\% as down-sampling was further intensified. This analysis, performed on hundreds of known pulsars within the survey, yielded consistent results (see Figure \ref{fig:downsampled_sn}). The observed decline in sensitivity arose from sub-optimal down-sampling procedures, effectively reducing 2-bit data to 1-bit precision (the scaling
    factor was incorrect by a factor 4$^{1/2}$=2) leading to an effective 1-bit sampling. It is 
    well known that in the presence of noise, 1-bit sampling leads to a reduction in signal-to-noise ratio by a factor of $\sqrt{1/\pi} \approx 0.79$, or approximately $1.25$ times. Therefore, the observed losses when down-sampling by a factor of 4 or more align with the characteristics of 1-bit sampling. The extent to which 1-bit sampling affected observations across the entirety of the HTRU-S LowLat survey remains uncertain. However, it is plausible that some pulsars with FFT $\rm S/N > 8.0$ may have been overlooked during candidate inspection, or that these candidates were simply not considered for further analysis.

    \item The significant increase in pulsar discoveries can be attributed, in no small part, to the refined candidate selection and folding techniques employed in our study. In the case of the original HTRU-S LowLat survey processing, both MSP and normal pulsar candidates were subjected to a constant false alarm threshold of 8, as assumed by \cite{cherry15} and \cite{cameron2020high}. This threshold was uniformly applied to both $\rm S/N_{FFT}$ and $\rm S/N_{fold}$. Additionally, candidates with up to 16 harmonic sums were selected for folding, aligning with the prevailing practice in many previous and ongoing pulsar searches. However, it's a known fact that using higher number of harmonics e.g., 32 can increase the search sensitivity for narrower duty cycle pulsars \citep[e.g.,][]{handbook04}. The analysis of simulated pulsars in \citet{sengar_23} has also provided a critical insight in to this. Due to the effects of incoherent harmonic summing, particularly for narrow duty cycle pulsars, the $\rm S/N_{fft}$ can often fall below the false-alarm threshold of 8. Upon folding, the $\rm S/N_{fold}$ can increase to levels up to 2.5 times higher. This underscores the importance of avoiding high false alarm $\rm S/N_{fft}$ thresholds in any survey. In retrospect, assuming accurate down-sampling was executed in prior survey processing phases, it is noteworthy that among the 71 pulsars, 30 exhibited $\rm S/N_{fft}$$<$8 (see Figure \ref{fig:pulsars_fft_fold_sn}). Intriguingly, out of these, 10 were detected with 32 harmonic sums, suggesting that the inclusion of higher harmonic sums played another helpful role in discovering approximately 10\% of the pulsars that might have otherwise been overlooked with at most a 16-harmonic sum processing.

    \item While less probable, there are additional factors that could account for missed pulsar detections. Notably, the previous segmented acceleration search pipeline generated an abundance of candidates — approximately 15 times more — leading to an elevated false-alarm threshold above S/N=8. Consequently, delving deeper into S/N levels revealed previously undiscovered pulsars, resulting in the detection of many pulsars well below FFT $\rm S/N = 8$. Furthermore, the presence of a known, particularly bright pulsar within an observation can obscure the detection of other, fainter pulsar signals. This occurs because the dominant signal from the bright pulsar overwhelms the candidate list with multiple harmonics, DMs, and accelerations, potentially relegating the weaker pulsar signal lower down the list.

\end{itemize}

\section{Discussion and Conclusion}
\label{sec:discussion_and_conclusion}

\begin{figure}
\centering
\includegraphics[width=0.45\textwidth]{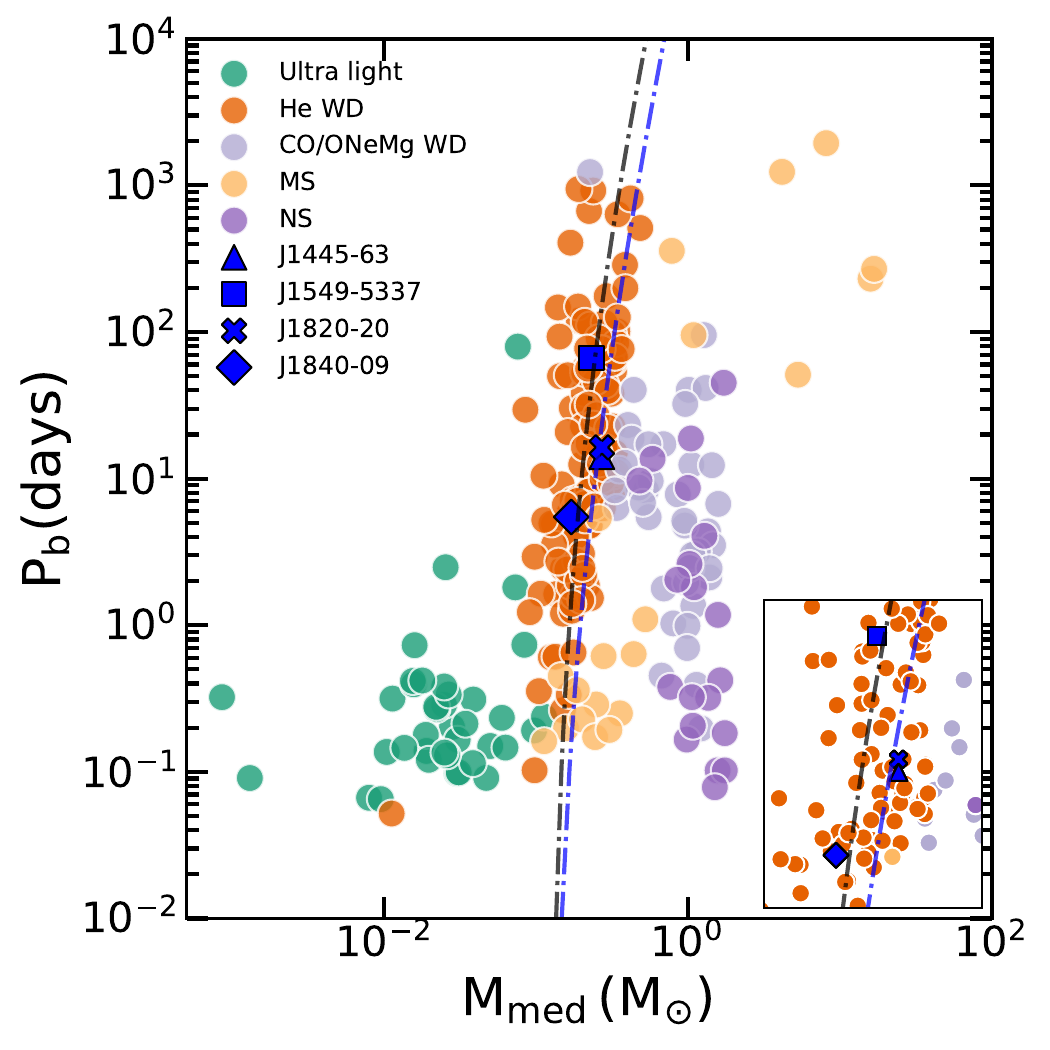}
\caption{The orbital period (in days) vs median companion masses of Galactic MSPs of different types in binary systems. The data for these systems were taken from the \texttt{PSRCAT}. The new binary MSPs presented in this work are labelled and shown in blue with different markers. The dashed dotted line in blue is the theoretical prediction obtained by \citet{tauris_99} for MSP-He WD systems formed as a final product of LMXBs, and the dashed-dotted line in black is the correlation for MSP-He WDs systems obtained by \citet{hui_18}.}

\label{fig:m_pb_relation}
\end{figure} 

We have presented a GPU accelerated search of $\sim 87 \%$ data of the HTRU-S LowLat pulsar survey, utilizing the full 72-minute observations at their native time resolution. This coherent acceleration search has led to the discovery of 71 new pulsars which have been reported for the first time in this study. Simultaneously, two other parallel reprocessing efforts also resulted in another 23 new pulsars (see Section \ref{sec:htru_survey_intro}), totalling 94 pulsar discoveries in these reprocessings endeavours. In particular, this reprocessing effort has significantly increased the new pulsars yield in the survey by $\sim 75\%$, making it arguably one of the most fruitful reprocessing to date in terms of the fraction of new pulsar discoveries as compared to pulsar discoveries in first processing of any survey (note that the PMPS underwent reprocessing 8 times to increase the pulsar yield by $\sim 50\%$). These findings can be attributed to two key factors. Firstly, the enhancement in sensitivity achieved by reprocessing the survey data without any decimation and to lesser extent a wider acceleration space. Secondly, the adoption of a technique involving 32 harmonic sums, along with a straightforward yet highly effective candidate sorting candidates from near the FFT noise floor, folding and classification techniques, as detailed in \citet{sengar_23}.

The use of GPUs in this work has significantly improved the processing speed of the survey and achieved performance nearly three orders of magnitude faster than CPU based codes (see Section \ref{sec:methods}). The newer versions of GPUs with increased parallel processing capabilities and higher memory bandwidth would further boost pulsar searching performances. The only bottleneck in our processing was folding of candidates as the folding was done on CPU based codes. Future pulsar searches, such as those conducted with the SKA are expected to generate a plethora of candidates, making it impractical to fold all of them. In this context, the development and implementation of GPU-based folding software becomes one of the top priorities in pulsar searches.

This reprocessing has effectively met all the primary goals of the survey. It has been successful in finding the predicted number of pulsars for the survey based on population synthesis. However, slight discrepancies remain, particularly in the case of MSP population, where the survey identified 42 pulsars, including new discoveries, as opposed to the expected 63 \citep{levin013}. Furthermore, the reprocessing has excelled in discovering a population of pulsars characterized by high dispersion measure (DM), scattering, and at greater distances, in contrast to the initial sample of 100 pulsars identified in the survey. While these newly found pulsars exhibit lower flux densities, their larger distances place them within the same luminosity distribution as the previously identified HTRU-S LowLat pulsars.

In addition to the discovery of the double neutron star system PSR J1325−-6253 \citep{sengar_22}, this study has also unveiled four other millisecond binary pulsars. A comparison of these binary discoveries with the Galactic plane population indicates consistency with the typical ones found in the Galaxy with He-WD companions (refer to Figure \ref{fig:m_pb_relation}). In terms of binary pulsar detections, both \citet{cherry15} and \citet{cameron2020high} reported a total of 39 binary pulsars (28 known and 11 new). Our reprocessing efforts led to the identification of 48 binary pulsars (32 known, 11 previous HTRU-LowLat, and 5 new). Notably, four known binary pulsars, namely PSRs J1811$-$2405, J1822$-$0848, J1837$-$0822, and J1840$-$0643, were previously undetected, despite their relatively high $S/N_{\rm fft}$. The reasons for their prior non-detections remains uncertain. 

Moving forward, our focus will be on developing precise timing solutions for the pulsars introduced in this study, with the exception of PSR J1325--6253, which has already been published. This endeavour will not only enhance our understanding of the individual characteristics of these pulsars but will also contribute to characterizing the newly discovered population as a whole. While a significant portion of the HTRU-LowLat survey's objectives has been met in this work, there is still room for improvement. Apart from the discovery of PSR J1325--6253, no relativistic short-orbit binary pulsar has been discovered in this work. One potential explanation for this could be a lack of a jerk search. This was highlighted by the discovery of the highly relativistic binary pulsar, PSR J1757--1854 \citep{cameron2020high}, which displayed about a 30\% decrease in signal-to-noise ratio (SNR) during full-length observation compared to its half-segment counterpart. As discussed in Section \ref{sec:why_so_missed}, the down-sampling in the original segmented acceleration search fell short of optimal conditions. There could be opportunities to discover faint highly-accelerated binary pulsars by re-implementing the segmented search with the techniques employed in this study.

\section*{Data Availability}

The HTRU-S LowLat data used in this work will be shared on reasonable request to the HTRU-S LowLat collaboration.

\section*{Acknowledgements}

The Parkes Observatory is part of the Australia Telescope National Facility (grid.421683.a) which is funded by the Australian Government for operation as a National Facility managed by the Commonwealth Scientific and Industrial Research Organisation (CSIRO). We acknowledge the Wiradjuri people as the traditional owners of the Observatory site. Observations to localize some pulsars used the FBFUSE and APSUSE computing clusters for data acquisition, storage and analysis. These clusters were funded and installed by the Max-Planck-Institut für Radioastronomie (MPIfR) and the Max-Planck-Gesellschaft. The MeerKAT telescope is operated by the South African Radio Astronomy Observatory, which is a facility of the National Research Foundation, an agency of the Department of Science and Innovation. The authors acknowledge support from the Australian Research Council (ARC) Centre of Excellence for Gravitational Wave Discovery (OzGrav), through project number CE170100004. RS was partially supported by the National Science Foundation (NSF) grant AST-1816904. SS is a recipient of an ARC Discovery Early Career Research Award (DE220100241). RMS acknowledges support through ARC Future Fellowship FT190100155. This research greatly benefited from the OzSTAR supercomputer at the Swinburne University of Technology, a facility supported by both Swinburne University of Technology and the National Collaborative Research Infrastructure Strategy (NCRIS). Without the storage of the archival data and computing resources on OzSTAR this work would have not been possible. We further acknowledge that these results are based upon work initially presented in \citet{sengar_thesis}. While the present paper incorporates moderate modifications, the core of the earlier doctoral work remains intact.




\bibliographystyle{mnras}
\bibliography{htru_XIX} 

\begin{thebibliography}{}
\makeatletter
\relax
\def\mn@urlcharsother{\let\do\@makeother \do\$\do\&\do\#\do\^\do\_\do\%\do\~}
\def\mn@doi{\begingroup\mn@urlcharsother \@ifnextchar [ {\mn@doi@} {\mn@doi@[]}}
\def\mn@doi@[#1]#2{\def\@tempa{#1}\ifx\@tempa\@empty \href {http://dx.doi.org/#2} {doi:#2}\else \href {http://dx.doi.org/#2} {#1}\fi \endgroup}
\def\mn@eprint#1#2{\mn@eprint@#1:#2::\@nil}
\def\mn@eprint@arXiv#1{\href {http://arxiv.org/abs/#1} {{\tt arXiv:#1}}}
\def\mn@eprint@dblp#1{\href {http://dblp.uni-trier.de/rec/bibtex/#1.xml} {dblp:#1}}
\def\mn@eprint@#1:#2:#3:#4\@nil{\def\@tempa {#1}\def\@tempb {#2}\def\@tempc {#3}\ifx \@tempc \@empty \let \@tempc \@tempb \let \@tempb \@tempa \fi \ifx \@tempb \@empty \def\@tempb {arXiv}\fi \@ifundefined {mn@eprint@\@tempb}{\@tempb:\@tempc}{\expandafter \expandafter \csname mn@eprint@\@tempb\endcsname \expandafter{\@tempc}}}

\bibitem[\protect\citeauthoryear{{Agazie} et~al.,}{{Agazie} et~al.}{2023}]{nanograv_21a}
{Agazie} G.,  et~al., 2023, \mn@doi [\apjl] {10.3847/2041-8213/acdac6}, \href {https://ui.adsabs.harvard.edu/abs/2023ApJ...951L...8A} {951, L8}

\bibitem[\protect\citeauthoryear{{Arzoumanian}, {Chernoff}  \& {Cordes}}{{Arzoumanian} et~al.}{2002}]{2002ApJ...568..289A}
{Arzoumanian} Z.,  {Chernoff} D.~F.,   {Cordes} J.~M.,  2002, \mn@doi [\apj] {10.1086/338805}, \href {https://ui.adsabs.harvard.edu/abs/2002ApJ...568..289A} {568, 289}

\bibitem[\protect\citeauthoryear{{Bailes} et~al.,}{{Bailes} et~al.}{2011}]{bailes_11}
{Bailes} M.,  et~al., 2011, \mn@doi [Science] {10.1126/science.1208890}, \href {https://ui.adsabs.harvard.edu/abs/2011Sci...333.1717B} {333, 1717}

\bibitem[\protect\citeauthoryear{{Barsdell}, {Bailes}, {Barnes}  \& {Fluke}}{{Barsdell} et~al.}{2012}]{barsdel12}
{Barsdell} B.~R.,  {Bailes} M.,  {Barnes} D.~G.,   {Fluke} C.~J.,  2012, \mn@doi [\mnras] {10.1111/j.1365-2966.2012.20622.x}, \href {https://ui.adsabs.harvard.edu/abs/2012MNRAS.422..379B} {422, 379}

\bibitem[\protect\citeauthoryear{{Bates}, {Lorimer}, {Rane}  \& {Swiggum}}{{Bates} et~al.}{2014}]{bates_14}
{Bates} S.~D.,  {Lorimer} D.~R.,  {Rane} A.,   {Swiggum} J.,  2014, \mn@doi [\mnras] {10.1093/mnras/stu157}, \href {https://ui.adsabs.harvard.edu/abs/2014MNRAS.439.2893B} {439, 2893}

\bibitem[\protect\citeauthoryear{{Bhat}, {Cordes}, {Camilo}, {Nice}  \& {Lorimer}}{{Bhat} et~al.}{2004}]{bhat_04}
{Bhat} N.~D.~R.,  {Cordes} J.~M.,  {Camilo} F.,  {Nice} D.~J.,   {Lorimer} D.~R.,  2004, \mn@doi [\apj] {10.1086/382680}, \href {https://ui.adsabs.harvard.edu/abs/2004ApJ...605..759B} {605, 759}

\bibitem[\protect\citeauthoryear{{Cameron} et~al.,}{{Cameron} et~al.}{2018}]{cameron_18}
{Cameron} A.~D.,  et~al., 2018, \mn@doi [\mnras] {10.1093/mnrasl/sly003}, \href {https://ui.adsabs.harvard.edu/abs/2018MNRAS.475L..57C} {475, L57}

\bibitem[\protect\citeauthoryear{{Cameron} et~al.,}{{Cameron} et~al.}{2020}]{cameron2020high}
{Cameron} A.~D.,  et~al., 2020, \mn@doi [\mnras] {10.1093/mnras/staa039}, \href {https://ui.adsabs.harvard.edu/abs/2020MNRAS.493.1063C} {493, 1063}

\bibitem[\protect\citeauthoryear{{Champion} et~al.,}{{Champion} et~al.}{2016}]{champion2016MNRAS}
{Champion} D.~J.,  et~al., 2016, \mn@doi [\mnras] {10.1093/mnrasl/slw069}, \href {https://ui.adsabs.harvard.edu/abs/2016MNRAS.460L..30C} {460, L30}

\bibitem[\protect\citeauthoryear{Cordes \& Lazio}{Cordes \& Lazio}{2002}]{ne2001}
Cordes J.~M.,  Lazio T. J.~W.,  2002, arXiv preprint astro-ph/0207156

\bibitem[\protect\citeauthoryear{{Cromartie} et~al.,}{{Cromartie} et~al.}{2020}]{2020NatAs...4...72C}
{Cromartie} H.~T.,  et~al., 2020, \mn@doi [Nature Astronomy] {10.1038/s41550-019-0880-2}, \href {https://ui.adsabs.harvard.edu/abs/2020NatAs...4...72C} {4, 72}

\bibitem[\protect\citeauthoryear{{Deller} et~al.,}{{Deller} et~al.}{2019}]{deller_19}
{Deller} A.~T.,  et~al., 2019, \mn@doi [\apj] {10.3847/1538-4357/ab11c7}, \href {https://ui.adsabs.harvard.edu/abs/2019ApJ...875..100D} {875, 100}

\bibitem[\protect\citeauthoryear{{Dirson}, {P{\'e}tri}  \& {Mitra}}{{Dirson} et~al.}{2022}]{2022A&A...667A..82D}
{Dirson} L.,  {P{\'e}tri} J.,   {Mitra} D.,  2022, \mn@doi [\aap] {10.1051/0004-6361/202243305}, \href {https://ui.adsabs.harvard.edu/abs/2022A&A...667A..82D} {667, A82}

\bibitem[\protect\citeauthoryear{{Eatough}, {Kramer}, {Lyne}  \& {Keith}}{{Eatough} et~al.}{2013}]{eatough13a}
{Eatough} R.~P.,  {Kramer} M.,  {Lyne} A.~G.,   {Keith} M.~J.,  2013, \mn@doi [\mnras] {10.1093/mnras/stt161}, \href {https://ui.adsabs.harvard.edu/abs/2013MNRAS.431..292E} {431, 292}

\bibitem[\protect\citeauthoryear{{Faulkner} et~al.,}{{Faulkner} et~al.}{2004}]{faulkner04}
{Faulkner} A.~J.,  et~al., 2004, \mn@doi [\mnras] {10.1111/j.1365-2966.2004.08310.x}, \href {https://ui.adsabs.harvard.edu/abs/2004MNRAS.355..147F} {355, 147}

\bibitem[\protect\citeauthoryear{{Frail}, {Goss}  \& {Whiteoak}}{{Frail} et~al.}{1994}]{1994ApJ...437..781F}
{Frail} D.~A.,  {Goss} W.~M.,   {Whiteoak} J.~B.~Z.,  1994, \mn@doi [\apj] {10.1086/175038}, \href {https://ui.adsabs.harvard.edu/abs/1994ApJ...437..781F} {437, 781}

\bibitem[\protect\citeauthoryear{{Fruchter} et~al.,}{{Fruchter} et~al.}{1990}]{fruchter_90}
{Fruchter} A.~S.,  et~al., 1990, \mn@doi [\apj] {10.1086/168502}, \href {https://ui.adsabs.harvard.edu/abs/1990ApJ...351..642F} {351, 642}

\bibitem[\protect\citeauthoryear{{Gitika} et~al.,}{{Gitika} et~al.}{2023}]{gitika_23}
{Gitika} P.,  et~al., 2023, \mn@doi [\mnras] {10.1093/mnras/stad2841}, \href {https://ui.adsabs.harvard.edu/abs/2023MNRAS.526.3370G} {526, 3370}

\bibitem[\protect\citeauthoryear{{Han}, {Manchester}  \& {Qiao}}{{Han} et~al.}{1999}]{han_99}
{Han} J.~L.,  {Manchester} R.~N.,   {Qiao} G.~J.,  1999, \mn@doi [\mnras] {10.1046/j.1365-8711.1999.02544.x}, \href {https://ui.adsabs.harvard.edu/abs/1999MNRAS.306..371H} {306, 371}

\bibitem[\protect\citeauthoryear{{Han} et~al.,}{{Han} et~al.}{2021}]{fast_gpps_21}
{Han} J.~L.,  et~al., 2021, \mn@doi [Research in Astronomy and Astrophysics] {10.1088/1674-4527/21/5/107}, \href {https://ui.adsabs.harvard.edu/abs/2021RAA....21..107H} {21, 107}

\bibitem[\protect\citeauthoryear{{Haslam}, {Salter}, {Stoffel}  \& {Wilson}}{{Haslam} et~al.}{1982}]{haslam82}
{Haslam} C.~G.~T.,  {Salter} C.~J.,  {Stoffel} H.,   {Wilson} W.~E.,  1982, \aaps, \href {https://ui.adsabs.harvard.edu/abs/1982A&AS...47....1H} {47, 1}

\bibitem[\protect\citeauthoryear{{Hewish}, {Bell}, {Pilkington}, {Scott}  \& {Collins}}{{Hewish} et~al.}{1968}]{hbp68}
{Hewish} A.,  {Bell} S.~J.,  {Pilkington} J.~D.~H.,  {Scott} P.~F.,   {Collins} R.~A.,  1968, \mn@doi [\nat] {10.1038/217709a0}, \href {http://adsabs.harvard.edu/abs/1968Natur.217..709H} {217, 709}

\bibitem[\protect\citeauthoryear{{Hobbs} et~al.,}{{Hobbs} et~al.}{2020}]{uwl_parkes}
{Hobbs} G.,  et~al., 2020, \mn@doi [\pasa] {10.1017/pasa.2020.2}, \href {https://ui.adsabs.harvard.edu/abs/2020PASA...37...12H} {37, e012}

\bibitem[\protect\citeauthoryear{{Hotan}, {van Straten}  \& {Manchester}}{{Hotan} et~al.}{2004}]{hotan_04}
{Hotan} A.~W.,  {van Straten} W.,   {Manchester} R.~N.,  2004, \mn@doi [\pasa] {10.1071/AS04022}, \href {https://ui.adsabs.harvard.edu/abs/2004PASA...21..302H} {21, 302}

\bibitem[\protect\citeauthoryear{{Hui}, {Wu}, {Han}, {Kong}  \& {Tam}}{{Hui} et~al.}{2018}]{hui_18}
{Hui} C.~Y.,  {Wu} K.,  {Han} Q.,  {Kong} A.~K.~H.,   {Tam} P.~H.~T.,  2018, \mn@doi [\apj] {10.3847/1538-4357/aad5ec}, \href {https://ui.adsabs.harvard.edu/abs/2018ApJ...864...30H} {864, 30}

\bibitem[\protect\citeauthoryear{{Johnston}, {Lyne}, {Manchester}, {Kniffen}, {D'Amico}, {Lim}  \& {Ashworth}}{{Johnston} et~al.}{1992}]{johnston92a}
{Johnston} S.,  {Lyne} A.~G.,  {Manchester} R.~N.,  {Kniffen} D.~A.,  {D'Amico} N.,  {Lim} J.,   {Ashworth} M.,  1992, \mn@doi [\mnras] {10.1093/mnras/255.3.401}, \href {https://ui.adsabs.harvard.edu/abs/1992MNRAS.255..401J} {255, 401}

\bibitem[\protect\citeauthoryear{Kaplan et~al.,}{Kaplan et~al.}{2019}]{kaplan019}
Kaplan D.~L.,  et~al., 2019, \mn@doi [The Astrophysical Journal] {10.3847/1538-4357/ab397f}, 884, 96

\bibitem[\protect\citeauthoryear{{Keane}, {Kramer}, {Lyne}, {Stappers}  \& {McLaughlin}}{{Keane} et~al.}{2011}]{keith011}
{Keane} E.~F.,  {Kramer} M.,  {Lyne} A.~G.,  {Stappers} B.~W.,   {McLaughlin} M.~A.,  2011, \mn@doi [\mnras] {10.1111/j.1365-2966.2011.18917.x}, \href {https://ui.adsabs.harvard.edu/abs/2011MNRAS.415.3065K} {415, 3065}

\bibitem[\protect\citeauthoryear{{Keith} et~al.,}{{Keith} et~al.}{2010}]{keith10}
{Keith} M.~J.,  et~al., 2010, \mn@doi [\mnras] {10.1111/j.1365-2966.2010.17325.x}, \href {https://ui.adsabs.harvard.edu/abs/2010MNRAS.409..619K} {409, 619}

\bibitem[\protect\citeauthoryear{{Knispel} et~al.,}{{Knispel} et~al.}{2013}]{knispel13}
{Knispel} B.,  et~al., 2013, \mn@doi [\apj] {10.1088/0004-637X/774/2/93}, \href {https://ui.adsabs.harvard.edu/abs/2013ApJ...774...93K} {774, 93}

\bibitem[\protect\citeauthoryear{{Kocz}, {Briggs}  \& {Reynolds}}{{Kocz} et~al.}{2010}]{kocz10}
{Kocz} J.,  {Briggs} F.~H.,   {Reynolds} J.,  2010, \mn@doi [\aj] {10.1088/0004-6256/140/6/2086}, \href {https://ui.adsabs.harvard.edu/abs/2010AJ....140.2086K} {140, 2086}

\bibitem[\protect\citeauthoryear{{Konar} \& {Deka}}{{Konar} \& {Deka}}{2019}]{2019JApA...40...42K}
{Konar} S.,  {Deka} U.,  2019, \mn@doi [Journal of Astrophysics and Astronomy] {10.1007/s12036-019-9608-z}, \href {https://ui.adsabs.harvard.edu/abs/2019JApA...40...42K} {40, 42}

\bibitem[\protect\citeauthoryear{{Kramer} et~al.,}{{Kramer} et~al.}{2021}]{kramer_21}
{Kramer} M.,  et~al., 2021, \mn@doi [Physical Review X] {10.1103/PhysRevX.11.041050}, \href {https://ui.adsabs.harvard.edu/abs/2021PhRvX..11d1050K} {11, 041050}

\bibitem[\protect\citeauthoryear{{Krishnakumar}, {Mitra}, {Naidu}, {Joshi}  \& {Manoharan}}{{Krishnakumar} et~al.}{2015}]{krishna15}
{Krishnakumar} M.~A.,  {Mitra} D.,  {Naidu} A.,  {Joshi} B.~C.,   {Manoharan} P.~K.,  2015, \mn@doi [\apj] {10.1088/0004-637X/804/1/23}, \href {https://ui.adsabs.harvard.edu/abs/2015ApJ...804...23K} {804, 23}

\bibitem[\protect\citeauthoryear{{Lentati}, {Kerr}, {Dai}, {Shannon}, {Hobbs}  \& {Os{\l}owski}}{{Lentati} et~al.}{2017}]{lenati_17}
{Lentati} L.,  {Kerr} M.,  {Dai} S.,  {Shannon} R.~M.,  {Hobbs} G.,   {Os{\l}owski} S.,  2017, \mn@doi [\mnras] {10.1093/mnras/stx580}, \href {https://ui.adsabs.harvard.edu/abs/2017MNRAS.468.1474L} {468, 1474}

\bibitem[\protect\citeauthoryear{{Levin} et~al.,}{{Levin} et~al.}{2010}]{levin10}
{Levin} L.,  et~al., 2010, \mn@doi [\apjl] {10.1088/2041-8205/721/1/L33}, \href {https://ui.adsabs.harvard.edu/abs/2010ApJ...721L..33L} {721, L33}

\bibitem[\protect\citeauthoryear{Levin et~al.,}{Levin et~al.}{2013}]{levin013}
Levin L.,  et~al., 2013, \mn@doi [\mnras] {10.1093/mnras/stt1103}, 434, 1387–1397

\bibitem[\protect\citeauthoryear{{Levin} et~al.,}{{Levin} et~al.}{2018}]{levin_18}
{Levin} L.,  et~al., 2018, in {Weltevrede} P.,  {Perera} B.~B.~P.,  {Preston} L.~L.,   {Sanidas} S.,  eds, ~ Vol. 337, Pulsar Astrophysics the Next Fifty Years. pp 171--174 (\mn@eprint {arXiv} {1712.01008}), \mn@doi{10.1017/S1743921317009528}

\bibitem[\protect\citeauthoryear{{Lorimer}}{{Lorimer}}{2011}]{psrpop_11}
{Lorimer} D.,  2011, {PSRPOP: Pulsar Population Modelling Programs}, Astrophysics Source Code Library, record ascl:1107.019 (\mn@eprint {ascl} {1107.019})

\bibitem[\protect\citeauthoryear{{Lorimer} \& {Kramer}}{{Lorimer} \& {Kramer}}{2004}]{handbook04}
{Lorimer} D.~R.,  {Kramer} M.,  2004, {Handbook of Pulsar Astronomy}

\bibitem[\protect\citeauthoryear{Lorimer et~al.,}{Lorimer et~al.}{2006}]{lorimer06}
Lorimer D.~R.,  et~al., 2006, \mn@doi [\mnras] {10.1111/j.1365-2966.2006.10887.x}, 372, 777–800

\bibitem[\protect\citeauthoryear{{Lorimer}, {Bailes}, {McLaughlin}, {Narkevic}  \& {Crawford}}{{Lorimer} et~al.}{2007}]{lorimer_07}
{Lorimer} D.~R.,  {Bailes} M.,  {McLaughlin} M.~A.,  {Narkevic} D.~J.,   {Crawford} F.,  2007, \mn@doi [Science] {10.1126/science.1147532}, \href {https://ui.adsabs.harvard.edu/abs/2007Sci...318..777L} {318, 777}

\bibitem[\protect\citeauthoryear{Lyne, Mankelow, Bell  \& Manchester}{Lyne et~al.}{2000}]{lyne2000}
Lyne A.~G.,  Mankelow S.~H.,  Bell J.~F.,   Manchester R.~N.,  2000, \mn@doi [\mnras] {10.1046/j.1365-8711.2000.03517.x}, 316, 491–493

\bibitem[\protect\citeauthoryear{{Lyne} et~al.,}{{Lyne} et~al.}{2017}]{lyne17}
{Lyne} A.~G.,  et~al., 2017, \mn@doi [\apj] {10.3847/1538-4357/834/1/72}, \href {https://ui.adsabs.harvard.edu/abs/2017ApJ...834...72L} {834, 72}

\bibitem[\protect\citeauthoryear{{Manchester} et~al.,}{{Manchester} et~al.}{2001}]{pmps01}
{Manchester} R.~N.,  et~al., 2001, \mn@doi [\mnras] {10.1046/j.1365-8711.2001.04751.x}, \href {https://ui.adsabs.harvard.edu/abs/2001MNRAS.328...17M} {328, 17}

\bibitem[\protect\citeauthoryear{{Manchester}, {Hobbs}, {Teoh}  \& {Hobbs}}{{Manchester} et~al.}{2005}]{psrcat05a}
{Manchester} R.~N.,  {Hobbs} G.~B.,  {Teoh} A.,   {Hobbs} M.,  2005, \mn@doi [\aj] {10.1086/428488}, \href {https://ui.adsabs.harvard.edu/abs/2005AJ....129.1993M} {129, 1993}

\bibitem[\protect\citeauthoryear{{Manchester}, {Fan}, {Lyne}, {Kaspi}  \& {Crawford}}{{Manchester} et~al.}{2006}]{manchester_06}
{Manchester} R.~N.,  {Fan} G.,  {Lyne} A.~G.,  {Kaspi} V.~M.,   {Crawford} F.,  2006, \mn@doi [\apj] {10.1086/505461}, \href {https://ui.adsabs.harvard.edu/abs/2006ApJ...649..235M} {649, 235}

\bibitem[\protect\citeauthoryear{{McLaughlin} et~al.,}{{McLaughlin} et~al.}{2006}]{mclaughlin06}
{McLaughlin} M.~A.,  et~al., 2006, \mn@doi [\nat] {10.1038/nature04440}, \href {https://ui.adsabs.harvard.edu/abs/2006Natur.439..817M} {439, 817}

\bibitem[\protect\citeauthoryear{{Mickaliger} et~al.,}{{Mickaliger} et~al.}{2012}]{mickaliger012}
{Mickaliger} M.~B.,  et~al., 2012, \mn@doi [\apj] {10.1088/0004-637X/759/2/127}, \href {https://ui.adsabs.harvard.edu/abs/2012ApJ...759..127M} {759, 127}

\bibitem[\protect\citeauthoryear{{Middleditch} \& {Kristian}}{{Middleditch} \& {Kristian}}{1984}]{middleditch_84}
{Middleditch} J.,  {Kristian} J.,  1984, \mn@doi [\apj] {10.1086/161876}, \href {https://ui.adsabs.harvard.edu/abs/1984ApJ...279..157M} {279, 157}

\bibitem[\protect\citeauthoryear{{Morello} et~al.,}{{Morello} et~al.}{2019}]{morello19}
{Morello} V.,  et~al., 2019, \mn@doi [\mnras] {10.1093/mnras/sty3328}, \href {https://ui.adsabs.harvard.edu/abs/2019MNRAS.483.3673M} {483, 3673}

\bibitem[\protect\citeauthoryear{{Ng} et~al.,}{{Ng} et~al.}{2015}]{cherry15}
{Ng} C.,  et~al., 2015, \mn@doi [\mnras] {10.1093/mnras/stv753}, \href {https://ui.adsabs.harvard.edu/abs/2015MNRAS.450.2922N} {450, 2922}

\bibitem[\protect\citeauthoryear{{Ocker}, {Cordes}  \& {Chatterjee}}{{Ocker} et~al.}{2020}]{ocker_20}
{Ocker} S.~K.,  {Cordes} J.~M.,   {Chatterjee} S.,  2020, \mn@doi [\apj] {10.3847/1538-4357/ab98f9}, \href {https://ui.adsabs.harvard.edu/abs/2020ApJ...897..124O} {897, 124}

\bibitem[\protect\citeauthoryear{{Padmanabh} et~al.,}{{Padmanabh} et~al.}{2023}]{padmanabh_23}
{Padmanabh} P.~V.,  et~al., 2023, \mn@doi [\mnras] {10.1093/mnras/stad1900}, \href {https://ui.adsabs.harvard.edu/abs/2023MNRAS.524.1291P} {524, 1291}

\bibitem[\protect\citeauthoryear{{Philippov}, {Timokhin}  \& {Spitkovsky}}{{Philippov} et~al.}{2020}]{2020PhRvL.124x5101P}
{Philippov} A.,  {Timokhin} A.,   {Spitkovsky} A.,  2020, \mn@doi [\prl] {10.1103/PhysRevLett.124.245101}, \href {https://ui.adsabs.harvard.edu/abs/2020PhRvL.124x5101P} {124, 245101}

\bibitem[\protect\citeauthoryear{{Ransom}, {Cordes}  \& {Eikenberry}}{{Ransom} et~al.}{2003}]{ransom_03}
{Ransom} S.~M.,  {Cordes} J.~M.,   {Eikenberry} S.~S.,  2003, \mn@doi [\apj] {10.1086/374806}, \href {https://ui.adsabs.harvard.edu/abs/2003ApJ...589..911R} {589, 911}

\bibitem[\protect\citeauthoryear{{Reardon} et~al.,}{{Reardon} et~al.}{2023}]{reardon_23}
{Reardon} D.~J.,  et~al., 2023, \mn@doi [\apjl] {10.3847/2041-8213/acdd02}, \href {https://ui.adsabs.harvard.edu/abs/2023ApJ...951L...6R} {951, L6}

\bibitem[\protect\citeauthoryear{{Sengar}}{{Sengar}}{2023}]{sengar_thesis}
{Sengar} R.,  2023, PhD thesis, Swinburne University of technology, \url {http://hdl.handle.net/1959.3/471917}

\bibitem[\protect\citeauthoryear{{Sengar} et~al.,}{{Sengar} et~al.}{2022}]{sengar_22}
{Sengar} R.,  et~al., 2022, \mn@doi [\mnras] {10.1093/mnras/stac821}, \href {https://ui.adsabs.harvard.edu/abs/2022MNRAS.512.5782S} {512, 5782}

\bibitem[\protect\citeauthoryear{{Sengar} et~al.,}{{Sengar} et~al.}{2023}]{sengar_23}
{Sengar} R.,  et~al., 2023, \mn@doi [\mnras] {10.1093/mnras/stad508}, \href {https://ui.adsabs.harvard.edu/abs/2023MNRAS.522.1071S} {522, 1071}

\bibitem[\protect\citeauthoryear{{Staelin}}{{Staelin}}{1969}]{staelin_69}
{Staelin} D.~H.,  1969, \mn@doi [IEEE Proceedings] {10.1109/PROC.1969.7051}, \href {https://ui.adsabs.harvard.edu/abs/1969IEEEP..57..724S} {57, 724}

\bibitem[\protect\citeauthoryear{{Stairs}}{{Stairs}}{2004}]{stairs_04}
{Stairs} I.~H.,  2004, \mn@doi [Science] {10.1126/science.1096986}, \href {https://ui.adsabs.harvard.edu/abs/2004Sci...304..547S} {304, 547}

\bibitem[\protect\citeauthoryear{{Stappers} \& {Kramer}}{{Stappers} \& {Kramer}}{2016}]{trapum_16}
{Stappers} B.,  {Kramer} M.,  2016, in MeerKAT Science: On the Pathway to the SKA. p.~9

\bibitem[\protect\citeauthoryear{{Staveley-Smith} et~al.,}{{Staveley-Smith} et~al.}{1996}]{multibeam96}
{Staveley-Smith} L.,  et~al., 1996, \pasa, \href {https://ui.adsabs.harvard.edu/abs/1996PASA...13..243S} {13, 243}

\bibitem[\protect\citeauthoryear{{Tan} et~al.,}{{Tan} et~al.}{2018}]{2018ApJ...866...54T}
{Tan} C.~M.,  et~al., 2018, \mn@doi [\apj] {10.3847/1538-4357/aade88}, \href {https://ui.adsabs.harvard.edu/abs/2018ApJ...866...54T} {866, 54}

\bibitem[\protect\citeauthoryear{{Tauris} \& {Savonije}}{{Tauris} \& {Savonije}}{1999}]{tauris_99}
{Tauris} T.~M.,  {Savonije} G.~J.,  1999, \mn@doi [\aap] {10.48550/arXiv.astro-ph/9909147}, \href {https://ui.adsabs.harvard.edu/abs/1999A&A...350..928T} {350, 928}

\bibitem[\protect\citeauthoryear{{Tauris} et~al.,}{{Tauris} et~al.}{2017}]{tauris17}
{Tauris} T.~M.,  et~al., 2017, \mn@doi [\apj] {10.3847/1538-4357/aa7e89}, \href {https://ui.adsabs.harvard.edu/abs/2017ApJ...846..170T} {846, 170}

\bibitem[\protect\citeauthoryear{{Wongphechauxsorn} et~al.,}{{Wongphechauxsorn} et~al.}{2024}]{jompoj_23}
{Wongphechauxsorn} J.,  et~al., 2024, \mn@doi [\mnras] {10.1093/mnras/stad3283}, \href {https://ui.adsabs.harvard.edu/abs/2024MNRAS.527.3208W} {527, 3208}

\bibitem[\protect\citeauthoryear{{Yao}, {Manchester}  \& {Wang}}{{Yao} et~al.}{2017}]{ymw16}
{Yao} J.~M.,  {Manchester} R.~N.,   {Wang} N.,  2017, \mn@doi [\apj] {10.3847/1538-4357/835/1/29}, \href {https://ui.adsabs.harvard.edu/abs/2017ApJ...835...29Y} {835, 29}

\bibitem[\protect\citeauthoryear{{Young}}{{Young}}{2011}]{2011yera.confE..47Y}
{Young} N.,  2011, in 41st Young European Radio Astronomers Conference. p.~47

\bibitem[\protect\citeauthoryear{{van den Heuvel} \& {Bonsema}}{{van den Heuvel} \& {Bonsema}}{1984}]{vanden_84}
{van den Heuvel} E.~P.~J.,  {Bonsema} P.~T.~J.,  1984, \aap, \href {https://ui.adsabs.harvard.edu/abs/1984A&A...139L..16V} {139, L16}

\makeatother
\end{thebibliography}


\bsp	
\label{lastpage}
\end{document}